**WebRTC Metadata and IP Leakage in Modern Browsers: A Cross-Platform Measurement Study**

Ahmed Fouad Kadhim KOYSHA [1][https://orcid.org/0009-0004-2967-724X], Aytug BOYACI [2*] [0000-0003-1016-3439], RAFET AKDENİZ [3] [0000-0003-3017-7651]

[1] Department of Computer Engineering, Faculty of Engineering, Istanbul Aydin University, Istanbul, Türkiye Correspondence: afouadkoysha@stu.aydin.edu.tr

[2] Department of Computer Engineering Air Force Academy Istanbul, Turkey aytugboyaci@firat.edu.tr

[3] Department of Electronics, University of Istanbul, rafetakdeniz@aydin.edu.tr

[*] Correspondence: Ahmed Fouad Kadhim KOYSHA (afouadkoysha@stu.aydin.edu.tr)

## Abstract

Web Real-Time Communication (WebRTC) enables real-time peer-to-peer communication, but its Interactive Connectivity Establishment (ICE) process can unintentionally expose internal and public IP addresses as metadata. This paper presents a cross-platform measurement study of WebRTC metadata leakage using current (2025) builds of Chrome, Brave, Firefox, and Tor on desktop and mobile platforms. Experiments were conducted across semi-trusted Wi-Fi and untrusted mobile carrier networks. Results show that Chrome remains the most leakage-prone, disclosing LAN or Carrier-Grade NAT (CGNAT) addresses on mobile and metadata on desktop; Brave avoids direct IP leaks but exposes session-stable mDNS identifiers; Firefox provides strong protection on desktop but leaks internal IPs on Android; and Tor consistently prevents all forms of leakage. We introduce a structured threat model for semi-trusted environments and evaluate the limitations of mDNS obfuscation. Finally, we propose layered mitigation strategies combining browser defaults, institutional safeguards, and user controls. Findings demonstrate that while direct LAN leakage is declining, emerging vectors such as mDNS and CGNAT create persistent privacy risks requiring protocol-level redesign and policy action.

## Ethics Statement

This study did not involve human or animal experimentation, nor did it collect any personally identifiable or sensitive information. The research was conducted as a technical assessment of the WebRTC protocol using open-source tools deployed in a controlled university Wi-Fi environment. No direct user interaction or active network interference occurred. All ethical standards were maintained during data collection and analysis to ensure that the privacy and security of individuals and institutions were not compromised.

## 1. Introduction

Web Real-Time Communication (WebRTC) enables direct peer-to-peer (P2P) communication through standardized JavaScript APIs, making it a cornerstone of modern browser-based applications such as teleconferencing, telemedicine, and enterprise collaboration. Its connectivity relies on the Interactive Connectivity Establishment (ICE) framework, which gathers potential connection paths using STUN and TURN servers. However, during this process, WebRTC may expose sensitive metadata—including internal and external IP addresses—even within encrypted browsing sessions. Unlike conventional vulnerabilities, this behavior is protocol-compliant, making it more difficult to detect or mitigate.

Prior research has highlighted this issue, often referred to as a "sanctioned metadata leak" (Al-Fannah, 2017). Fakis et al. (2020) and Feher et al. (2018) showed that browsers leaked both LAN and public IPs by default, while Tang et al. (2023) demonstrated how such leakage persists even with VPN tunneling. Yet, the persistence of WebRTC leakage across modern browser builds (2025) and platforms, particularly within semi-trusted environments, remains underexplored. Semi-trusted environments (e.g., university or enterprise Wi-Fi) differ from home or public networks: users are authenticated (via student or enterprise credentials), but devices remain unmanaged under bring-your-own-device (BYOD) policies. This creates a unique attack surface for passive adversaries capable of profiling users or conducting reconnaissance without direct interaction.

This study does not claim the discovery of a novel vulnerability. Instead, it provides an updated and reproducible measurement of a well-known but underexamined privacy risk, addressing gaps in prior literature where most experiments relied on outdated builds or controlled lab conditions, without systematic evaluation of unmanaged BYOD networks. By testing up-to-date versions of Chrome, Brave, Firefox, and Tor across both Windows 11 and Android, we deliver a contemporary perspective on how leakage patterns differ between Chromium-based and non-Chromium browsers, and how recent mitigations—such as mDNS obfuscation—perform in practice.

Against this backdrop, the contributions of this study are fourfold:

1. Cross-platform evaluation of internal IP leakage across four major browsers on both desktop and mobile, using reproducible test scripts.
2. Structured threat model for semi-trusted BYOD environments, clarifying the unique risks of internal IP and metadata exposure.
3. Empirical analysis of mDNS obfuscation, showing partial effectiveness but residual fingerprinting risks.

4. Actionable mitigation strategies, combining browser-level policies, institutional safeguards, and end-user defenses.

By situating our analysis within real-world semi-trusted environments and evaluating modern browser builds, this study provides a timely update that bridges empirical evidence with practical mitigation strategies—offering insights relevant to browser developers, network administrators, and end users alike.

## 1.2 Motivation and Research Scope

Web Real-Time Communication (WebRTC) has become a core component of modern web applications, enabling real-time peer-to-peer (P2P) communication directly in the browser without additional plugins. While this capability has revolutionized areas such as teleconferencing, remote education, and enterprise collaboration, it also introduces a subtle yet critical privacy risk: the silent exposure of internal and external IP addresses during Interactive Connectivity Establishment (ICE) negotiations. Unlike conventional vulnerabilities, this behavior is protocol-compliant and often invisible to both end users and security tools. The metadata disclosed through ICE candidates can be exploited for browser fingerprinting, user tracking, and even internal network reconnaissance.

The motivation for this study stems from the need to provide an updated and reproducible assessment of these risks under modern conditions. Despite WebRTC's widespread adoption, its privacy implications—particularly in semi-trusted environments such as university campuses and enterprise Wi-Fi networks—remain underexplored. These networks typically enforce authentication but permit unmanaged bring-your-own-device (BYOD) access, creating a unique threat surface where passive adversaries can silently harvest metadata through in-browser scripts.

Although browser vendors have introduced mitigations such as mDNS hostname obfuscation, questions remain regarding their effectiveness and consistency across platforms. Many prior studies have focused on legacy builds, VPN contexts, or controlled testbeds, offering limited insight into today's real-world deployments. This work narrows its scope to metadata exposure and structured threat modeling, deliberately excluding performance or quality-of-service aspects. By testing current browser builds (2025) across Windows 11 and Android, our framework offers a lightweight, reproducible, browser-native methodology that reflects realistic attack scenarios. To our knowledge, this is among the first systematic evaluations of mDNS obfuscation across both desktop and mobile platforms in real-world network environments.

## 1.3 Research Contribution and Questions

While WebRTC metadata leakage has been documented in prior work, few studies have systematically measured its persistence across modern browsers, mobile platforms, and semi-trusted BYOD environments. Earlier evaluations emphasized outdated versions, simulated testbeds, or VPN-protected scenarios. In contrast, this paper contributes an updated, cross-platform measurement study that highlights inconsistencies in mitigation strategies and documents the residual risks of mDNS-based obfuscation.

The contributions of this study are fourfold:

1. Cross-platform evaluation of internal IP leakage across four major browsers (Chrome, Brave, Firefox, and Tor) on both desktop (Windows 11) and mobile (Android), conducted in two phases: (i) semi-trusted university Wi-Fi and (ii) residential/workplace BYOD networks.
2. Structured threat model tailored for semi-trusted environments, clarifying the unique risks of internal IP disclosure under BYOD access.
3. Empirical analysis of mDNS obfuscation, revealing its partial effectiveness and the continued risks of device fingerprinting.
4. Lightweight and reproducible testing framework, implemented in standard JavaScript, with actionable mitigation strategies for browser developers, administrators, and end users.

To guide our methodology and analysis, we defined the following research questions:

- RQ1: Does internal IP leakage persist in modern desktop and mobile browsers, and under what conditions?
- RQ2: How do leakage patterns vary across browser families (Chromium vs. non-Chromium) and operating systems (Windows vs. Android)?
- RQ3: What are the privacy implications of such leakage in semi-trusted environments, and how can adversaries exploit exposed metadata?
- RQ4: How effective is mDNS obfuscation in reducing IP exposure, and what residual risks remain for fingerprinting or profiling?
- RQ5: Which technical and policy-level mitigations can reduce or eliminate metadata exposure without significantly degrading WebRTC functionality?

Rather than presenting a new vulnerability, this work offers an up-to-date, reproducible measurement study of a known but underexamined privacy risk. By situating our findings within structured threat environments, we provide empirical evidence that informs actionable recommendations and supports independent validation through open datasets and a proof-of-concept design.

## 2. Literature Review

### 2.1 WebRTC and IP Address Leakage

Recent research has highlighted significant vulnerabilities in WebRTC technology, particularly with respect to unintentional IP address exposure. Fakis et al. (2020) conducted an empirical evaluation of popular web browsers Chrome, Firefox, and Edge and demonstrated that even when VPNs or private browsing modes are enabled, real IP addresses may still be leaked. Chrome was found to be especially prone to this issue, as WebRTC bypasses protective layers and exposes sensitive network metadata through ICE candidates. The authors stressed that users relying solely on browser-based privacy features might be unaware of this exposure. Manual deactivation of WebRTC or the use of privacy-enhancing extensions was recommended to address this threat.

Similarly, Feher et al. (2018) provided a detailed analysis of WebRTC's potential to leak internal and public IP addresses, even in semi-trusted environments like academic institutions. The study utilized JavaScript-based proof-of-concept code to demonstrate metadata harvesting from browsers using ICE candidates. This emphasized that WebRTC's architecture, while functionally efficient, poses serious risks to user anonymity and privacy.

Unlike Fakis et al. (2020) and Tang et al. (2023), our evaluation explicitly excludes VPN-protected sessions, focusing instead on baseline browser behavior in unmanaged BYOD environments. This ensures that findings reflect default browser behavior without interference from third-party tunneling.

### 2.2 Encryption Trade-offs and Quality of Service

The impact of encryption protocols on WebRTC performance has been a focal point of multiple studies. Moreno et al. (2021) examined how enabling secure protocols such as DTLS and TLS affect key QoS indicators including call establishment time, jitter, and maximum packet latency. Their findings showed mixed effects: in some cases, encryption introduced latency, while in others—particularly in Wi-Fi environments—it improved jitter stability due to buffering.

Tang et al. (2023) also noted performance-security trade-offs in peer-to-peer streaming networks. Their study revealed that WebRTC can leak user IP addresses to unauthorized peers, even under VPN and proxy protection. This occurs because WebRTC's STUN server connections are made outside the VPN tunnel, exposing metadata directly.

### 2.3 Metadata Leakage and Browser Behavior

Further investigations have explored how WebRTC behaves as a conduit for metadata leakage. Zhao (2024) introduced FProbe, a static analysis tool that detects fingerprinting activities in JavaScript code. The study found that many websites embed scripts capable of identifying users based on screen configuration, plugin data, and ICE negotiation parameters, suggesting that browser-based protocols like WebRTC can facilitate stealthy surveillance.

Other researchers have pointed out how poor configuration of STUN servers or weak signaling protocols further amplify the leakage risk (PANORAMIX Project, 2020; Raza et al., 2024). These studies emphasized the need for secure implementations rather than sole reliance on protocol-level specifications such as RFC 8826.

### 2.4 Proposed Secure Architectures

To mitigate these threats, several architecture-level solutions have been proposed. Raza et al. (2024) introduced a trust-based WebRTC framework that incorporates federated identities and attribute-based access control (ABAC) to prevent unauthorized peer connections. Tian and Jiang (2024) designed a dual-layer encryption mechanism where DTLS-SRTP encrypts media traffic while an application-layer protocol secures metadata. This system achieved both low latency and high media quality, demonstrating the feasibility of secure-by-design WebRTC systems.

Deshmukh et al. (2023) contributed a practical implementation of a WebRTC-based video conferencing system with features like peer-to-peer media exchange, NAT traversal using STUN/TURN, and secure transmission via DTLS-SRTP. While the system showed high performance in real-world tests, it lacked mechanisms to address metadata leakage and signaling vulnerabilities, highlighting areas for improvement that our study aims to address.

### 2.5 Summary of Prior Work

Across the surveyed literature, two themes consistently emerge. First, WebRTC remains prone to metadata leakage, with internal and external IP addresses being exposed under a variety of deployment conditions (Fakis et al., 2020; Feher et al., 2018). Second, although encryption and protocol-level safeguards (DTLS, TLS, SRTP) improve confidentiality of media streams, they do not adequately mitigate metadata disclosure, leaving users vulnerable to tracking and profiling (Tang et al., 2023; Moreno et al., 2021).

At the same time, research efforts have explored stronger architectural defenses, ranging from federated trust frameworks (Raza et al., 2024) to dual-layer encryption schemes (Tian & Jiang, 2024). While promising, these proposals often remain conceptual or are evaluated in lab-like settings rather than deployed networks. Furthermore, studies that did attempt practical browser testing often focused on legacy versions or narrow conditions (e.g., VPN-enabled sessions), limiting their relevance for current real-world environments.A comparative overview of the most relevant studies is provided in Table 1.

Table 1. Summary of Prior Studies on WebRTC Security and Privacy

| Study | Focus | Methodology / Tools | Key Findings | Gap / Limitation |
|---|---|---|---|---|
| Fakis et al. (2020) | IP leakage despite VPNs | Browser-based empirical tests (Chrome, Firefox, Edge) | IP leakage persists, esp. in Chrome | Focused on VPN contexts, outdated versions |
| Feher et al. (2018) | Leakage in semi-trusted (academic) networks | JS PoC, ICE candidate harvesting | Silent IP metadata leaks | Limited to desktop, older builds |
| Tang et al. (2023) | Metadata leaks under VPN/proxy | Protocol-level analysis, STUN configs | STUN bypasses VPN → critical exposure | Did not assess mobile / modern versions |
| Moreno et al. (2021) | QoS trade-offs under encryption | Experimental performance metrics (DTLS/TLS) | Encryption affects latency/jitter | Not focused on leakage/privacy |
| Zhao (2024) | Fingerprinting via ICE params | Static JS analysis (FProbe tool) | ~0.78% of sites use WebRTC fingerprinting | No direct leakage measurement |
| Raza et al. (2024) | Trust-based secure WebRTC | Federated identity + ABAC framework | Improved access control | Conceptual, not deployed/tested |
| Tian & Jiang (2024) | Metadata + media encryption | Dual-layer DTLS-SRTP + app-layer protocol | Low latency + strong security | Not evaluated in browser-native contexts |
| Deshmukh et al. (2023) | Scalable WebRTC conferencing | Real-world P2P systemw/STUN/TURN | High perf., metadata still exposed | No mitigation of ICE leakage |
| PANORAMIX (2020) | Privacy in messaging platforms | Case studies | Protocol design flaws weaken anonymity | Narrow scope, not browsers |
| Pardhi & Sonsare (2023) | Secure-by-design WebRTC | Private session initiation, E2E encryption | Metadata risks reduced | Prototype, not mainstream browsers |
| Sultan et al. (2023) | ML for abuse detection | Streaming data analysis | Detected hijacks, abuse patterns | Focus on misuse, not leakage |
| Mahmoud & Abozariba (2025) | WebRTC for IoT/edge | Systematic review | Scalability + privacy concerns | Theoretical, not browser-based |

### 2.6 Research Gap and Positioning

Despite a decade of awareness around WebRTC metadata exposure, three significant gaps remain:

1. Outdated Testing Baselines: Most prior experiments relied on legacy builds or desktop-only configurations. Few studies systematically evaluate leakage across modern (2025) browser releases on both desktop and mobile platforms.
2. Limited Contextual Environments: While semi-trusted networks (e.g., BYOD university Wi-Fi) represent realistic and increasingly common settings, they remain underexplored compared to home or public Wi-Fi scenarios.
3. Partial Mitigation Evaluation: Existing literature acknowledges mDNS obfuscation but rarely measures its practical effectiveness or residual risks, such as fingerprinting based on hostname patterns.

This study directly addresses these gaps by:

- Providing an updated cross-platform measurement of WebRTC IP leakage across Chrome, Brave, Firefox, and Tor on Windows 11 and Android.
- Introducing the first empirical analysis of mDNS obfuscation in practice, assessing both its strengths and limitations. To our knowledge, this is among the first studies to evaluate mDNS obfuscation across both desktop and mobile platforms in real-world networks.
- Situating the results within a structured threat model tailored to semi-trusted BYOD environments, where authenticated but unmanaged devices create unique attack surfaces.
- Delivering a lightweight, reproducible JavaScript-based testing framework that enables independent validation without custom browser builds.

By filling these gaps, this study positions itself as a timely update that complements prior VPN- and lab-focused research with real-world, browser-native experiments in unmanaged BYOD contexts.

Taken together, the reviewed literature demonstrates both the persistence of WebRTC metadata exposure and the lack of rigorous, cross-platform testing in semi-trusted BYOD environments. By addressing these gaps with reproducible, browser-native experiments, our study contributes timely empirical evidence that complements prior conceptual and VPN-focused research. In summary, our study contributes novelty not by disclosing a new vulnerability, but by updating the empirical evidence base with reproducible, browser-native experiments across modern desktop and mobile platforms. By incorporating mDNS obfuscation analysis and situating results in unmanaged BYOD networks, it offers both practical insights and a timely benchmark for future protocol and policy design.

## 3. Research Methods

This section outlines the methodology adopted to evaluate WebRTC metadata leakage across browsers, platforms, and network trust environments. The study was guided by five research questions:

- RQ1. Does internal IP leakage persist in modern desktop and mobile browsers, and under what conditions?
- RQ2. How do leakage patterns vary across browser families (Chromium vs. non-Chromium) and operating systems (Windows vs. Android)?
- RQ3. What are the privacy implications of such leakage in semi-trusted BYOD environments, and how can adversaries exploit exposed metadata?
- RQ4. How effective is mDNS obfuscation in reducing IP exposure, and what residual risks remain for fingerprinting or profiling?
- RQ5. Which technical and policy-level mitigations can reduce metadata exposure without significantly degrading WebRTC functionality?

To ensure ecological validity, the evaluation proceeded in three phases:

- Phase I (University Wi-Fi): Proof-of-concept measurements in a semi-trusted BYOD environment.
- Phase I-b (External VPS Deployment): Validation through HTTPS-compliant logging accessible from both university and home Wi-Fi.
- Phase II (Extended Networks): Broader evaluation across workplace Wi-Fi, residential Wi-Fi, and 4G ISP networks.

### 3.1 Script Preparation

A JavaScript-based probing tool was developed to evaluate WebRTC's default behavior during ICE candidate generation. The script initiated an RTCPeerConnection with a public STUN server (stun:stun.l.google.com:19302), which forced candidate collection.

Metadata extracted included:

- Internal and external IP addresses
- Browser user-agent and client hints (UA/UA-CH)
- Operating system and platform
- Browser language setting

- Device screen resolution and device pixel ratio (DPR)
- Timezone information

The extracted data was sent to a logging server via HTTP POST for structured storage and offline analysis see figure 1.

Algorithm 1. Client-side ICE metadata extraction (Phase I)
(Input: Browser with WebRTC enabled → Output: InternalIP, PublicIP, Metadata)

1. Initialize STUN server
2. Create RTCPeerConnection
3. Add data channel
4. Create offer and set local description
5. On ICE candidate event:
    - Extract IP via RegEx
    - If private IPv4 → record InternalIP
    - Else → record PublicIP
6. Collect metadata (UA, client hints, resolution, DPR, timezone, language)
7. Send payload to server via HTTP POST

### 3.2 Network Trust Models

To capture how WebRTC leakage varies under different assumptions of device management and authentication, we classified environments into three trust models:

- Trusted (Home Wi-Fi): Private ISP router, WPA2-Personal.
- Semi-Trusted (University/Workplace BYOD Wi-Fi): Authenticated Wi-Fi, unmanaged devices, shared NAT.
- Untrusted (Mobile/4G ISP): Carrier-grade NAT with minimal authentication, potential ISP-level logging.

These trust models align with prior classifications of network environments (e.g., Fakis et al., 2020; Mahmoud & Abozariba, 2025), ensuring that our methodology remains comparable to existing literature while extending the scope to unmanaged BYOD and mobile CGNAT scenarios.

### 3.3 Phase I: University Network Setup

To evaluate the practical feasibility of metadata leakage via WebRTC, a manual proof-of-concept deployment was conducted within a large-scale university network. The experiment aimed to simulate a realistic peer-to-peer (P2P) connection scenario and assess IP address exposure using default browser settings.

The JavaScript probe, identical in logic to that outlined in Algorithm 1, was executed manually on a university-issued device connected to the institutional Wi-Fi network via student credentials. The device accessed the network using a valid Student ID over the official university wireless infrastructure.

The script was injected directly into the browser Developer Tools Console without requiring any external software installation or user interaction beyond page loading. As illustrated in Figure 1, this approach ensured a covert, client-side-only interaction. No user prompts or visible indicators alerted the user to the metadata extraction process, simulating a stealth-based attack model.

This setup reflects a plausible real-world threat model where unsuspecting users in semi-trusted environments such as university or enterprise networks may be exposed to passive surveillance simply by loading a seemingly benign web page.

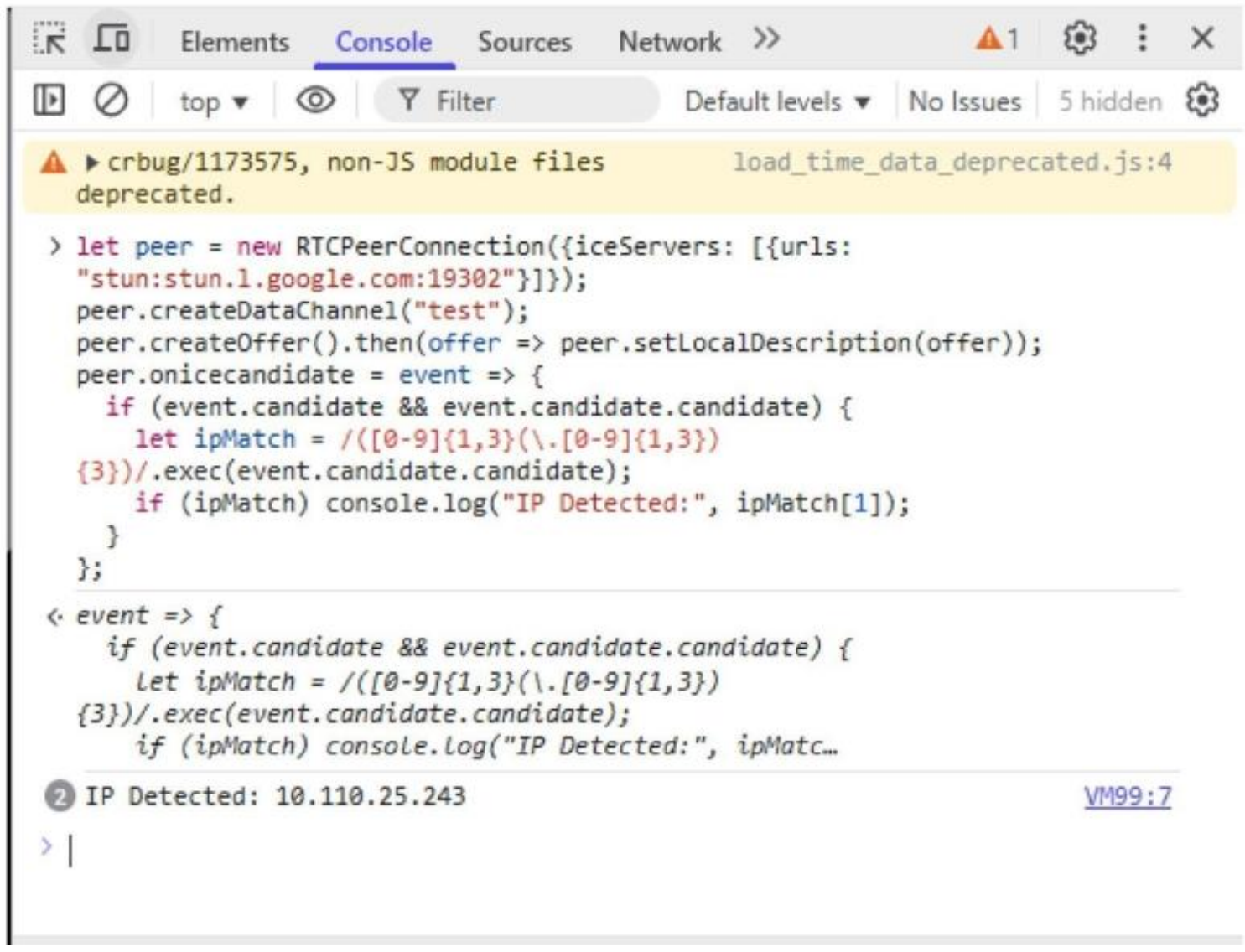

Figure 1. Metadata logs generated from ICE candidates during Phase I manual testing.

### 3.4 Phase I-b: External Domain Deployment (University + Home)

To simulate a realistic adversary scenario, the probe was deployed on an external VPS with HTTPS compliance. A domain (https://webrtc.serajtv.stream) was registered and configured with:

- steal.html — client-side probe page
- collect.php — server-side logging endpoint

The VPS was secured using Cloudflare DNS, TLS certificates, and Nginx reverse proxy. When accessed from university and home Wi-Fi networks, the probe silently collected internal and external IPs, as well as metadata (OS, UA, timezone, language) see figure 2.

Algorithm 2. Server-side Logging of WebRTC Metadata

```
IF HTTP request == POST THEN
  Read raw POST data
  Define log path (log.txt)
  Get current timestamp
  Append [timestamp + data] to log.txt
  Respond 200 ("Saved!")
ELSE
  Respond 405 ("Method Not Allowed")
END IF
```

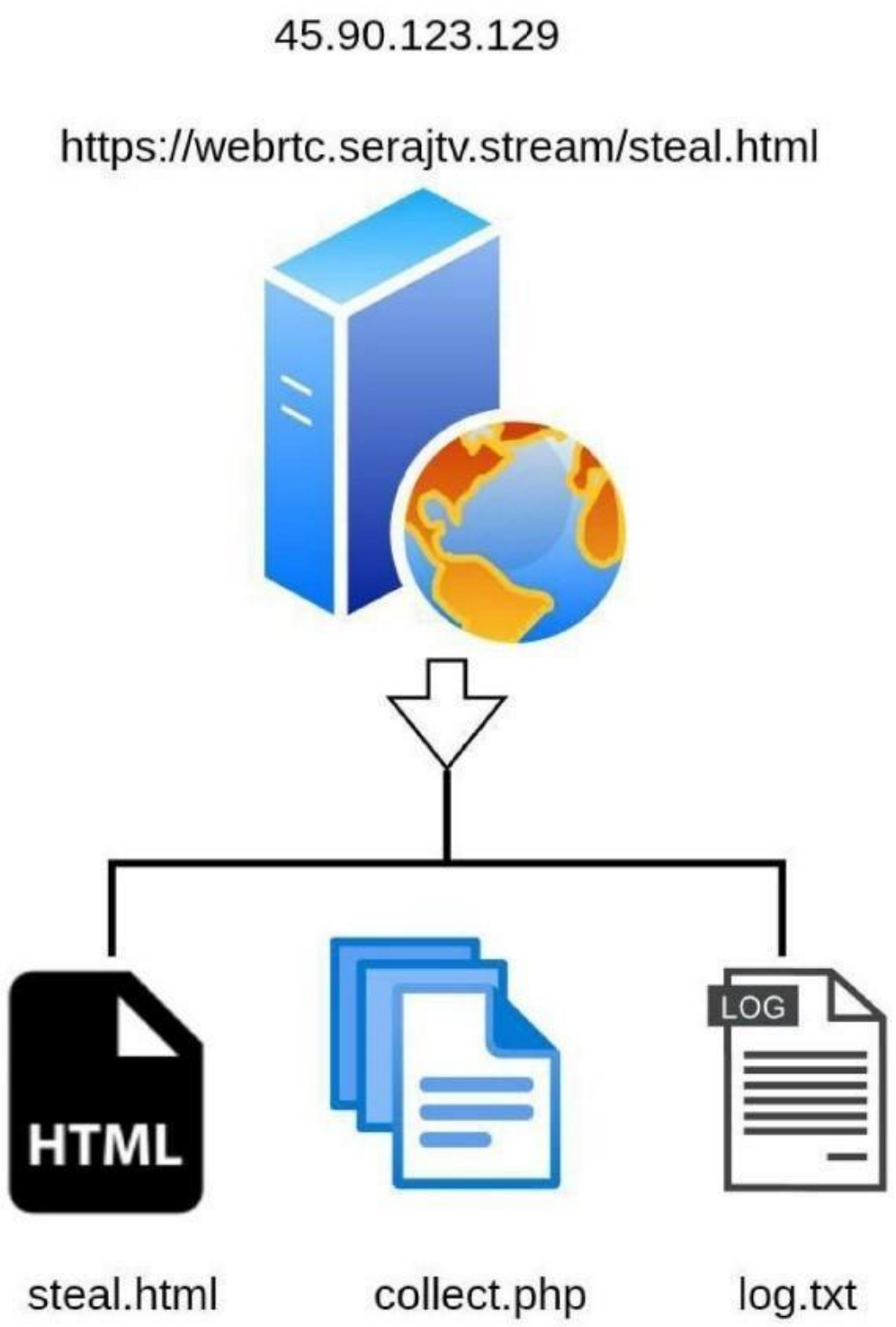

Figure 2. VPS file structure with deployed HTML probe and logging scripts.

3.5 Execution Flow and Real-Time Results (Phase I & Phase I-b Findings)

The proof-of-concept was executed in both a semi-trusted university Wi-Fi and a residential home Wi-Fi to validate WebRTC's ability to expose internal and external IP addresses without requiring user interaction or consent.

When a user visited the crafted URL (https://webrtc.serajtv.stream/steal.html), the browser automatically initiated a WebRTC handshake, triggering the Interactive Connectivity Establishment (ICE) process. As part of this negotiation, the browser generated ICE candidates that exposed:

- Internal IP addresses (e.g., 192.168.x.x or 10.x.x.x)
- External/NAT-translated public IPs (allocated by the ISP)

In addition, the probe collected browser-level metadata useful for passive fingerprinting:

- Browser type and version
- Operating system details
- Language and time zone
- Screen resolution and device pixel ratio (DPR, in Phase II)

This process required no explicit user interaction, no permission prompts, and no firewall alerts. Leakage was especially visible in legacy builds of Chrome and Firefox that lacked Multicast DNS (mDNS) obfuscation, a countermeasure later introduced to mask LAN IPs.

The end-to-end flow can be summarized as:

1. Peer A (victim) visits the crafted URL.
2. Embedded JavaScript probe executes silently.
3. A RTCPeerConnection is created with a public STUN server (stun.l.google.com:19302).
4. ICE candidates are gathered and parsed with RegEx.
5. Internal/External IPs + system metadata are encoded into JSON.
6. A POST request transmits the payload to the server logging endpoint (collect.php).

The server-side script (collect.php) logged results with timestamps into structured files, enabling retrospective validation and forensic analysis.

This workflow is illustrated in Figure 3, showing how ICE candidates and metadata were exfiltrated to the attacker-controlled VPS.

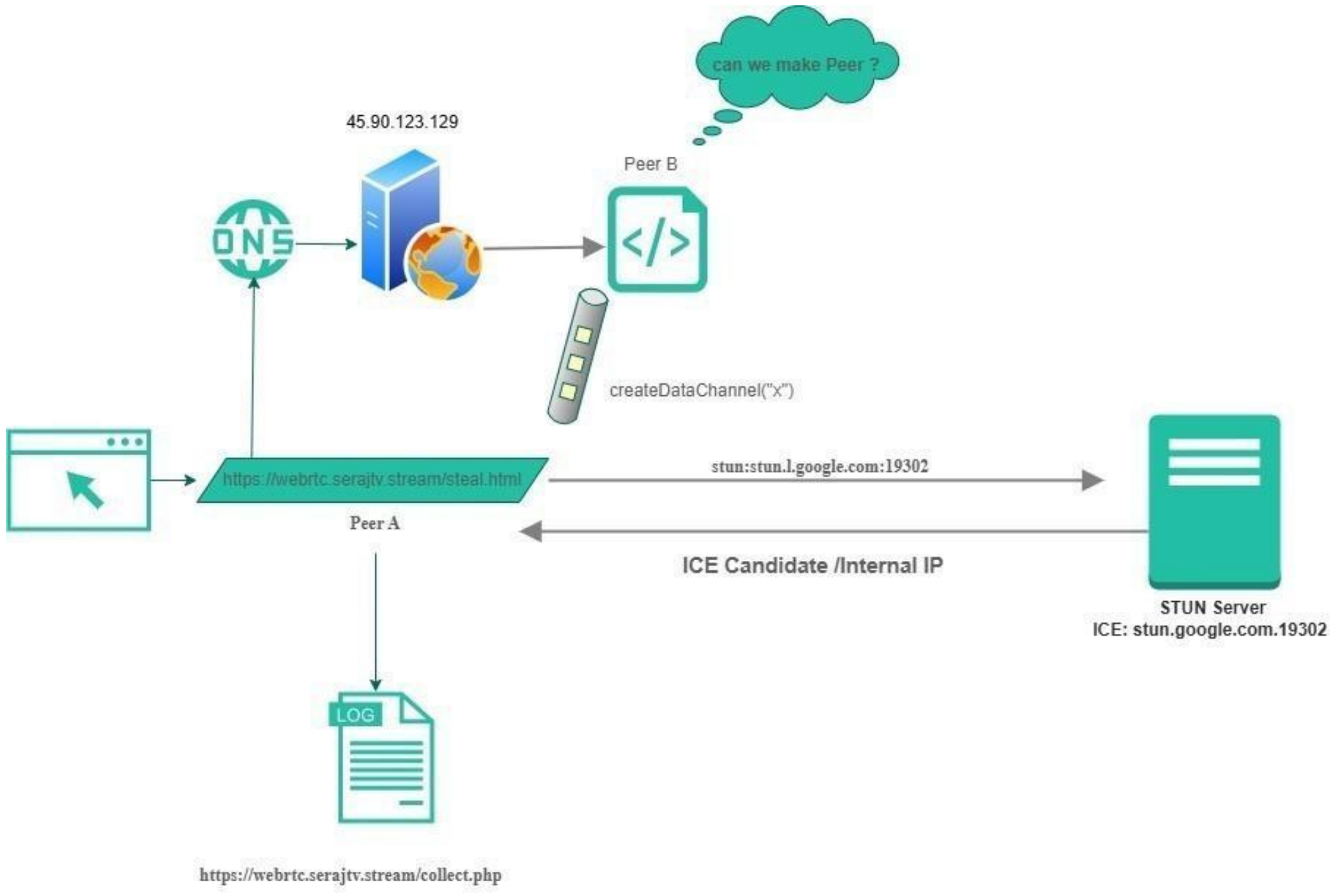

Figure 3. Phase I probe execution and metadata exfiltration workflow.

These initial findings established a baseline of IP and metadata leakage, later contrasted with Phase II results (see Section 5).

### 3.6 Reproducibility & Version Control

- Each test was repeated 3 times per device/browser/network.
- Caches and local storage were cleared, and browsers restarted.
- Only successful ICE collections were preserved.
- Logs were stored in JSON with metadata (browser version, OS, IP assignment, timestamp).

### 3.7 Phase II: Extended Networks

Phase II expanded testing to workplace Wi-Fi, home Wi-Fi, and mobile 4G ISP networks using Windows 11 (Pro/Home), Kali Linux 2025.2, and Galaxy S23 (Android 14). The probe was upgraded with richer metadata capture and integrity validation.

Algorithm 3 represents the novel extension of our probing framework, integrating additional dimensions of metadata such as mDNS hostname capture, UA-Client Hints, screen information, and SHA-256 integrity validation. These enhancements were not part of Phase I and were specifically introduced in Phase II to provide a more comprehensive and tamper-resistant measurement of browser leakage behaviors.

Algorithm 3. Enhanced Metadata Probe (Phase II)
*(Input: Browser with WebRTC enabled → Output: InternalIP, PublicIP, Metadata {UA, UA-CH, screenInfo, DPR, mDNS, TZ, Lang})*

1. Collect UA + UA-CH client hints
2. Gather screen resolution and DPR
3. Fetch public IP via external API
4. Initialize RTCPeerConnection with STUN
5. On ICE candidate event:
   - If private IPv4 → record InternalIP
   - If .local string → record mDNS hostname
6. Assemble JSON payload with timestamp + metadata
7. Compute SHA-256 hash of payload
8. POST payload to server logging endpoint

### 3.7 Devices and Builds

Table 2. Devices and browser builds used in testing

| ID | Device / OS | Network | Browser | Version / Build |
|---|---|---|---|---|
| D1 | Windows 11 Pro 24H2 | Work Wi-Fi | Chrome | 139.0.7258.139 (64-bit) |
| D2 | Windows 11 Pro 24H2 | Work Wi-Fi | Brave | 139.0.0.0 (64-bit) |
| D3 | Windows 11 Pro 24H2 | Work Wi-Fi | Firefox | 142.0 (64-bit) |
| D4 | Windows 11 Pro 24H2 | Work Wi-Fi | Tor Browser | 13.5.1 (Firefox ESR 128 base) |
| D5 | Kali Linux 2025.2 | Work Wi-Fi | Chromium | 137.0.7151.119 (64-bit) |
| D6 | Kali Linux 2025.2 | Work Wi-Fi | Brave | 1.80.115 (Chromium 138.0.7204) |
| D7 | Kali Linux 2025.2 | Work Wi-Fi | Firefox ESR | 128.11.0 (64-bit) |
| D8 | Kali Linux 2025.2 | Work Wi-Fi | Tor Browser | 14.5.5 (Firefox 128 base) |
| M1 | Galaxy S23 / Android 14 | Work Wi-Fi | Chrome Mobile | 139.0.7258.143 |
| M2 | Galaxy S23 / Android 14 | Work Wi-Fi | Brave Mobile | 1.81.135 (Chromium 139) |
| M3 | Galaxy S23 / Android 14 | Work Wi-Fi | Firefox Mobile | 130.0 |
| M4 | Galaxy S23 / Android 14 | Work Wi-Fi | Tor Mobile | 14.5.5 (128.13 ESR base) |
| M5 | Galaxy S23 / Android 14 | 4G ISP | Chrome Mobile | 139.0.7258.143 |
| M6 | Galaxy S23 / Android 14 | 4G ISP | Brave Mobile | 1.81.135 (Chromium 139) |
| M7 | Galaxy S23 / Android 14 | 4G ISP | Firefox Mobile | 130.0 |
| M8 | Galaxy S23 / Android 14 | 4G ISP | Tor Mobile | 14.5.5 (128.13 ESR base) |
| H1 | Windows 11 Home | Home Wi-Fi | Chrome | (v136.0-139.0.7258.139 (64-bit) |
| H2 | Windows 11 Home | Home Wi-Fi | Brave | 1.81.136 (Chromium 139.0.7258) |
| H3 | Windows 11 Home | Home Wi-Fi | Firefox | 142.0 (64-bit) |

### 3.8 Execution Flow

When users accessed the crafted URL, browsers automatically triggered ICE negotiations, exposing:

- Internal IPs (10.x.x.x / 192.168.x.x)

External/Public IPs (NAT-assigned by ISP)

- mDNS hostnames (*.local)
- Metadata: UA, UA-CH, OS, resolution, timezone

Phase I Findings: Chrome and Firefox Mobile leaked internal IPs; Brave leaked external only; Tor blocked leakage. Figure 4. Schematic visualization of WebRTC metadata flow and browser leakage patterns across environments. The diagram maps the journey from JavaScript ICE candidate retrieval to final metadata exposure, highlighting differences among Chrome, Firefox, Brave, and Tor.
Phase II Findings: Chrome still leaked internal IPs on home Wi-Fi/4G; Brave leaked mDNS; Firefox desktop obfuscated but mobile leaked; Tor blocked all.

This reproducible pipeline, progressing from manual console injection (Phase I) to VPS-based automated collection (Phase II), provided the foundation for the experimental setup described in Section 4.

## 4. Methodology and Experimental Setup

This section outlines the experimental setup designed to evaluate metadata exposure risks introduced by WebRTC in common browsing environments. Building on the methods described in Section 3, the setup operationalized both Phase I (baseline proof-of-concept in university and home Wi-Fi) and Phase II (extended networks: workplace BYOD, residential, and mobile 4G). The objective was to assess whether internal or external IP addresses and related metadata could be exposed through standard browser behavior, without requiring user interaction or privileged configurations.

### 4.1 Network Environments

Experiments were conducted across three representative network contexts, aligned with the trust models defined in Section 3.2:

- Semi-Trusted (University / Workplace Wi-Fi): Shared NAT firewalls, authenticated BYOD access.
- Trusted (Home Wi-Fi): ISP-provided router (WPA2-Personal) with dynamic IP allocation.
- Untrusted (Mobile 4G ISP): Carrier-grade NAT (CGNAT) with minimal authentication, exposing subscriber-unique private ranges (10.x.x.x).

These environments ensured coverage of realistic browsing scenarios encountered by end-users.

### 4.2 Devices and Browsers

Tests included both desktop and mobile platforms (Table 2, Section 3.7). Representative devices were:

- Desktop / Laptop: Windows 11 Pro (24H2), Windows 11 Home (24H2), Kali Linux 2025.2.
- Mobile: Samsung Galaxy S23 (Android 14).

Browsers tested:

- Google Chrome (v136.0 – v139.0)
- Mozilla Firefox (v139.0 – v142.0; ESR 128.11)
- Brave (v1.79 – v1.81, Chromium base)
- Tor Browser (v13.5 – v14.5, Firefox ESR 128 base)

All browsers were tested under default configurations, without extensions, VPNs, or privacy add-ons, ensuring results reflected out-of-the-box privacy protections.

### 4.3 WebRTC Probe Execution

Two complementary probing methods were deployed:

1. Phase I (Local Injection): Vanilla JavaScript embedded in a local HTML file and executed via browser Developer Tools Console. This leveraged the RTCPeerConnection API with a public STUN server (stun:stun.l.google.com:19302) to extract ICE candidates see Figure 4. Captured metadata included:
    - Internal IP (LAN: 192.168.x.x / 10.x.x.x)
    - External IP (ISP-assigned WAN)
    - OS, User-Agent, and Time Zone
    - Timestamp of connection attempt

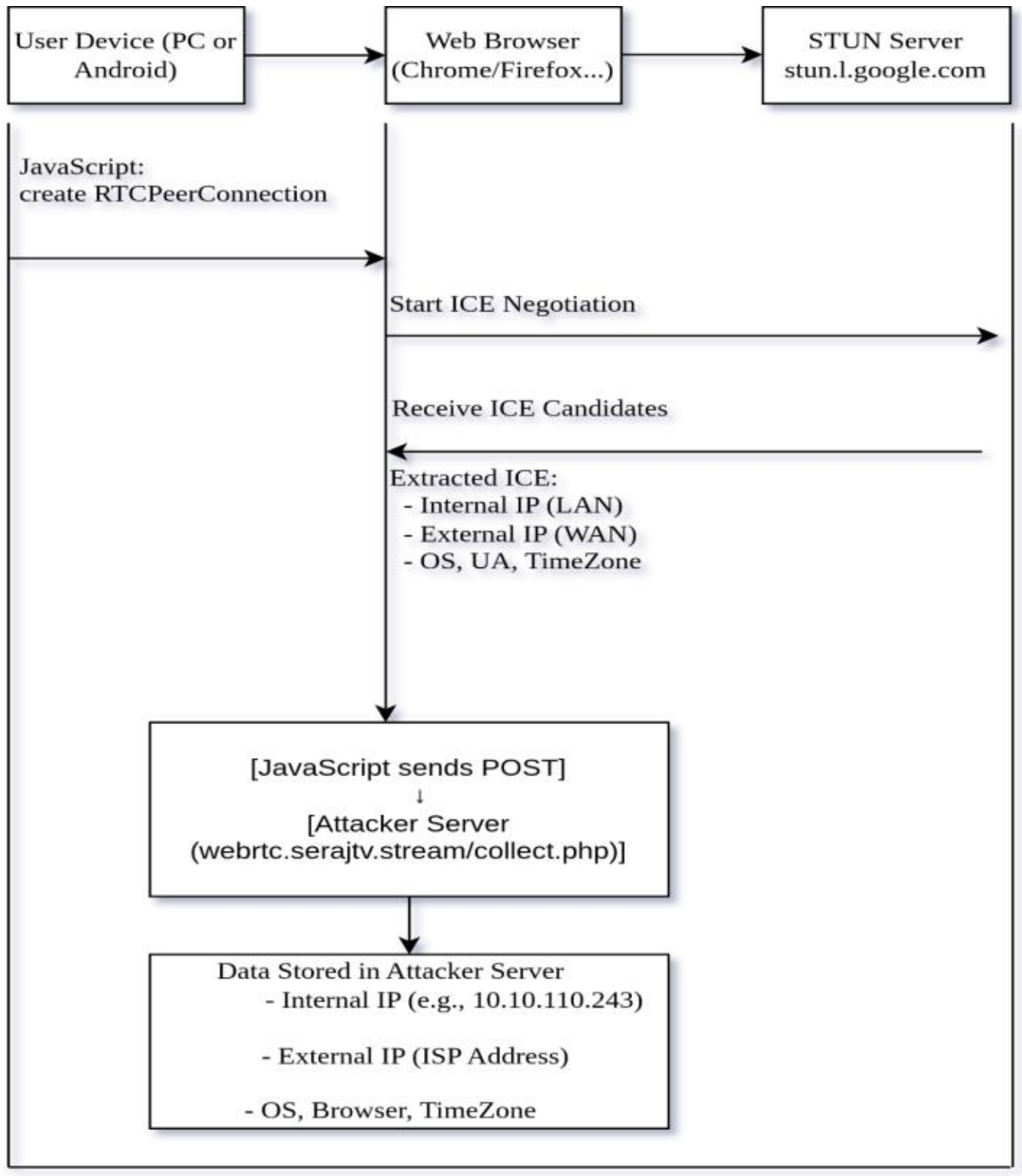

Figure 4. WebRTC Metadata Exposure Flow Diagram

2. Phase I-b & II (External VPS Deployment): To simulate a realistic adversary, the same script was hosted on an HTTPS-enabled VPS (webrtc.serajtv.stream) fronted by Cloudflare and Nginx reverse proxy. The client probe (steal.html) executed automatically when accessed, sending results via HTTP POST to a logging endpoint (collect.php).

The upgraded Phase II probe incorporated additional metadata collection:

- mDNS hostnames (introduced by Chrome/Brave/Firefox for obfuscation)
- UA-Client Hints (UA-CH) (architecture, platform version, bitness)
- Screen information & DPR (Device Pixel Ratio)
- SHA-256 integrity hashes of JSON payloads for tamper detection

This setup builds upon the logic described in Algorithm 1 (client-side probe), Algorithm 2 (server-side logging), and Algorithm 3 (enhanced Phase II probe).This workflow is illustrated in Figure 5.

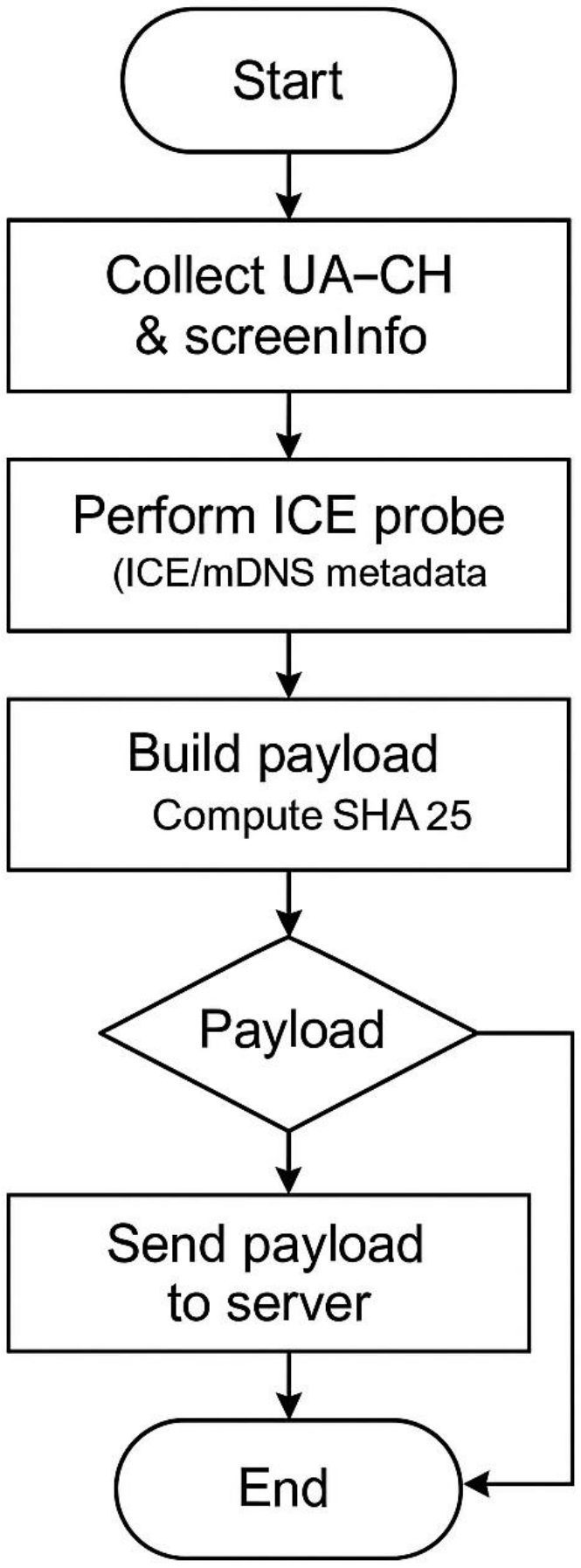

Figure 5. End-to-end execution pipeline

### 4.4 Test Procedures

To ensure reproducibility and comparability across environments:

- Each browser/device/network combination was tested three times.
- Caches and local storage were cleared, and browsers restarted before each run.
- Internal IPs were validated with ipconfig (Windows) and Android developer tools.
- Public IPs were verified with DNS utilities and cross-checked via dnschecker.org.
- Only successful ICE collections (with valid candidates + server logs) were preserved in the dataset.

This reproducibility protocol provided consistency across trials and ensured that results reflected actual browser defaults rather than cached behaviors.

### 4.5 Experimental Objectives

The setup was designed to directly address RQ1–RQ2:

- RQ1: Determine whether internal IP leakage persists in 2025 browser builds across desktop and mobile.
- RQ2: Compare leakage vectors (LAN IPs, CGNAT, mDNS, UA-CH) across platforms and trust environments.

By combining controlled console injection (Phase I) with external VPS logging (Phase I-b & II), the methodology reflects both lab-style validation and real-world adversary feasibility.

Subsequent sections (Results & Discussion) extend these findings to address RQ3–RQ5, analyzing privacy implications, mDNS effectiveness, and mitigation strategies.

This experimental setup established the basis for a systematic evaluation of leakage behaviors across browsers, platforms, and networks. The results of these experiments are presented in Section 5 (Results and Discussion).

## 5. Results

This section presents the results of our experimental evaluation of WebRTC metadata leakage across browsers, devices, and network environments. Findings are organized by behavior type and test phase. Results are presented with direct cross-references to the research questions (RQ1–RQ5) introduced in Section 1.

### 5.1 IP Leakage Observations (Phase I)

Across all scenarios, internal and external IP addresses were retrieved directly from ICE candidates via standard JavaScript execution, without user interaction or elevated privileges. In both university and home networks, Chrome-based browsers exposed internal and external IPs by default.

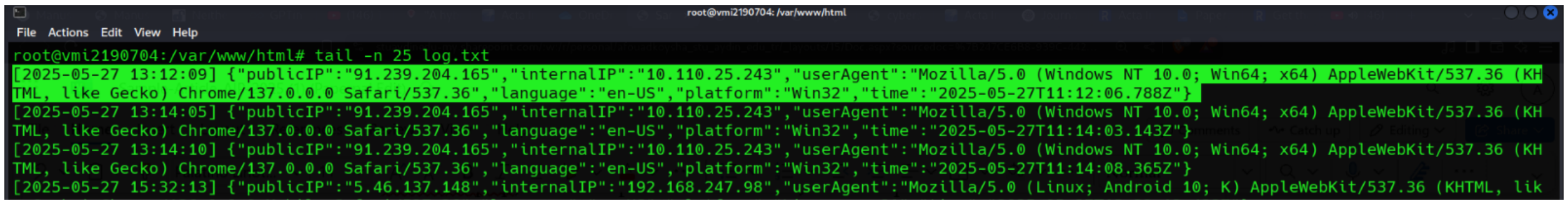

Figure 6. Chrome (Windows): WebRTC console log showing concurrent LAN + WAN exposure with system metadata.

Note on Browser Versioning. During Phase I, Chrome showed version 137.0.0.0 in candidate logs (developer-channel identifier), while the stable build used in analysis was v136.0.7103.114. Both exhibited identical leakage behavior; hence the discrepancy does not affect validity.

Internal IPs obtained via WebRTC matched the system IPs from ipconfig.

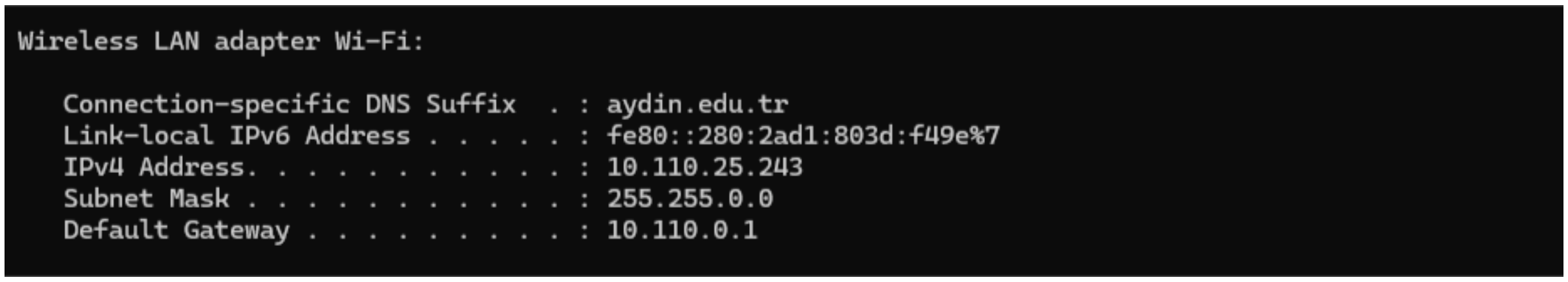

Figure 7. Windows ipconfig: console output showing internal IP matching WebRTC leak.

These initial observations confirm significant IP leakage across common browsers and networks, motivating a deeper comparison across browsers and devices.

### 5.1.1 Browser Comparisons

Browser behavior varied in the extent and type of metadata leaked via WebRTC:

- Google Chrome (v136.0.7103.114): disclosed *both* internal and external IPs plus system/browser metadata (OS, UA, time zone).
- Mozilla Firefox (v139.0.4): exposed external IP only; internal IPs masked by default.
- Brave (v1.79.123): similar to Firefox; internal IPs concealed, external IP exposed.
- Tor Browser: no identifiable leakage (neither IPs nor browser-level metadata).

Note on System Identification. The test machine ran Windows 11 (64-bit), yet the User-Agent reported Windows NT 10.0; Win64; x64 (compatibility string). This nominal difference does not affect ICE leakage mechanisms.

As illustrated below, Firefox logs contained external IP only during ICE negotiation.

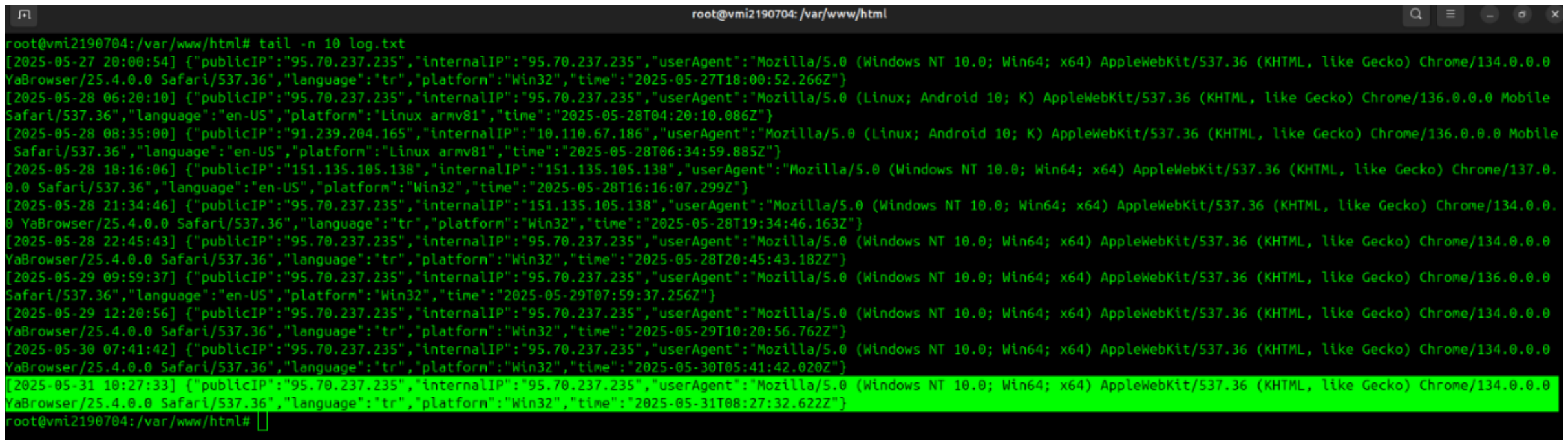

Figure 8. Firefox (Windows): console log showing external IP only during ICE negotiation.

Geolocation corroborated these differences: Chrome resolved to the user’s true location (Istanbul), while Tor routed via anonymized relays (Amsterdam).

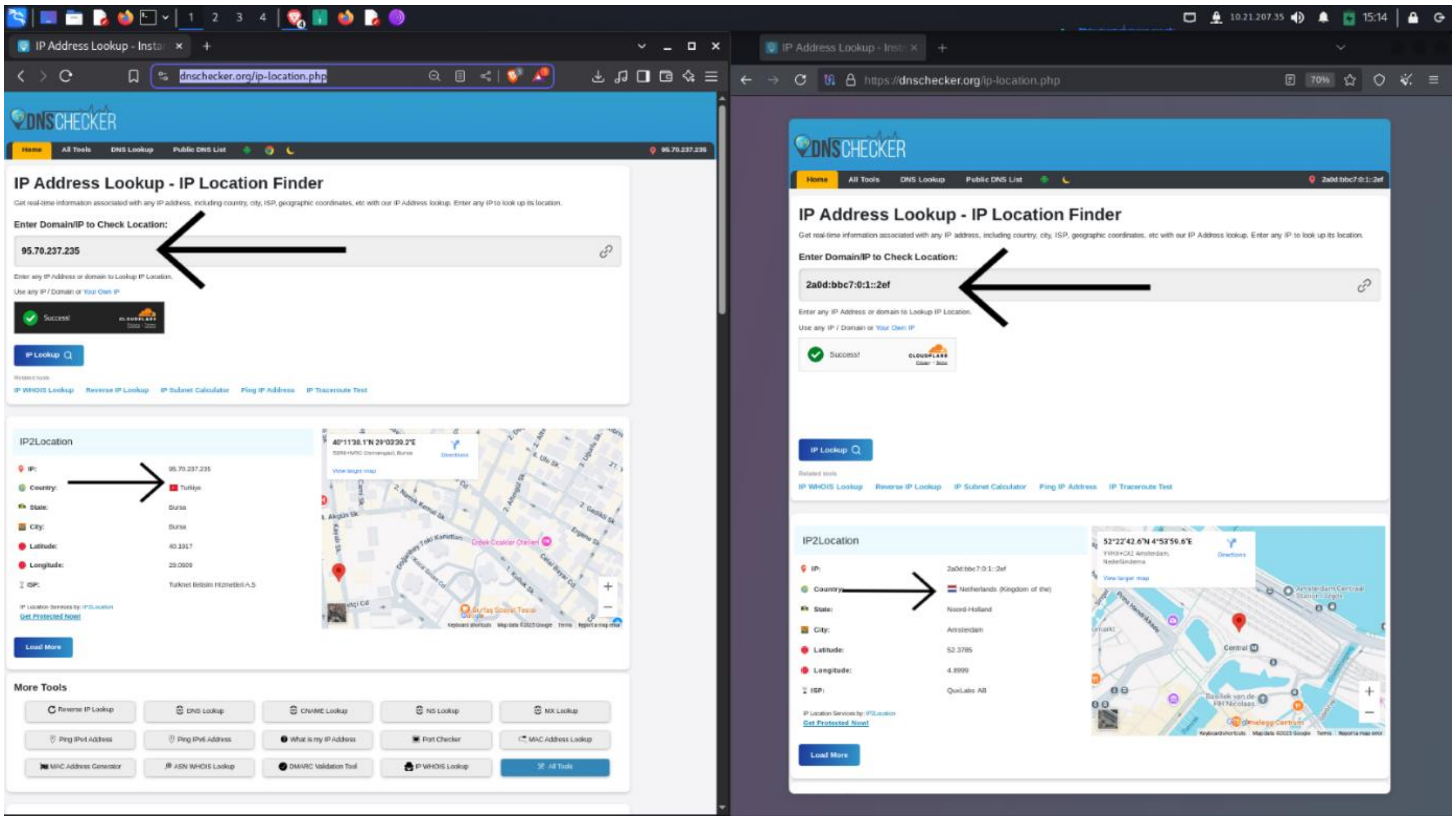

Figure 9. Geolocation results (dnschecker.org): Chrome resolves to Istanbul vs. Tor relays in Amsterdam.

Interpretation. Chrome presents the weakest privacy posture by combining LAN+WAN exposure and rich metadata. Firefox/Brave offer stronger defaults (no LAN), yet still reveal WAN. Tor fully suppresses leakage.

### 5.1.2 Mobile vs. PC Behavior

Android Chrome mirrored desktop behavior: LAN and WAN were exposed on page load without interaction. Tor on Android preserved full anonymity, consistent with desktop.

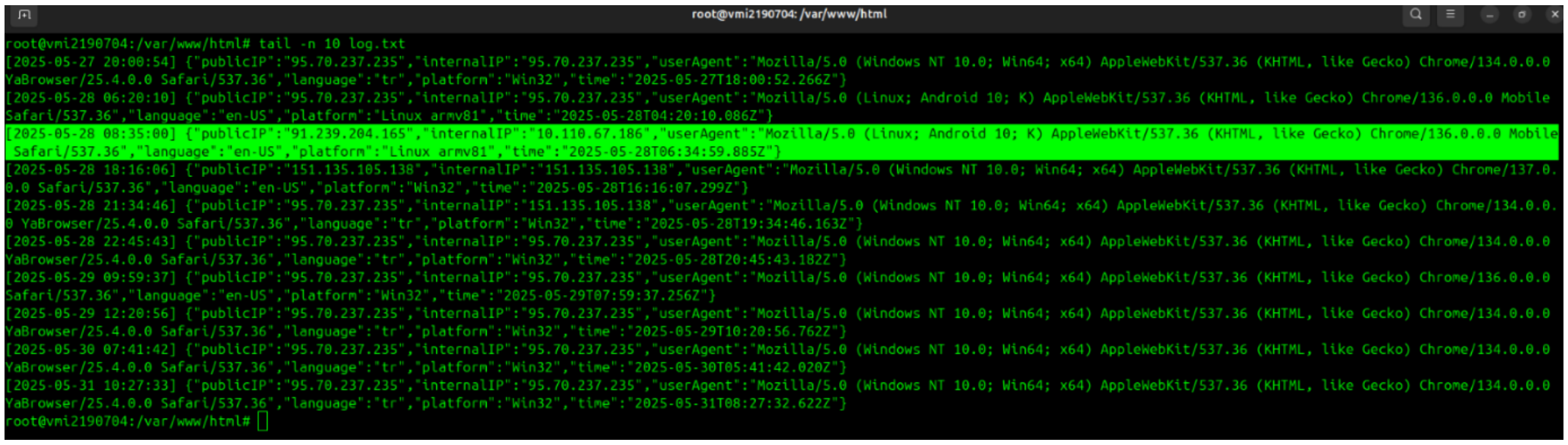

Figure 10. Chrome (Android): console log showing LAN + WAN exposure on page load.

Insight. Mobile builds (particularly Android) lag behind desktop in privacy-preserving defaults. This gap is likely due to differences in sandboxing models (mobile browsers rely more on shared system components), broader exposure of User-Agent Client Hints (UA-CH) that reveal device model and screen parameters, and vendor privacy policies that historically prioritize compatibility over strict isolation. As a result, Android platforms introduce a weaker privacy baseline compared to their desktop counterparts.

#### 5.1.3 Geolocation Tests

Using external geolocation services, Chrome's public IP accurately resolved to Istanbul, while Tor resolved to Amsterdam relays (see Figure 9). Third-party services can thus infer location from default ICE exposure, underscoring practical privacy risk.

#### 5.1.4 Summary of Findings

Table 3. Summary of WebRTC IP and Metadata Leakage Across Browsers and Devices.

| Device Type | Browser (Version) | Internal IP | External IP | Metadata (OS/UA/TZ) |
|---|---|---|---|---|
| PC | Chrome (136.0.7103) | Yes ☑ | Yes ☑ | Yes ☑ |
| PC | Firefox (139.0.4, 64-bit) | No ✕ | Yes ☑ | Yes ☑ |
| PC | Brave (1.79.123, 64-bit) | No ✕ | Yes ☑ | Yes ☑ |
| PC | Tor Browser | No ✕ | No ✕ | No ✕ |
| Android | Chrome (137.0.7151) | Yes ☑ | Yes ☑ | Yes ☑ |
| Android | Tor Browser | No ✕ | No ✕ | No ✕ |

These Phase I results establish the comparative baseline for modern-build testing in Phase II.

### 5.2 Phase II — Metadata Leakage in Modern Builds (2025)

Building on Phase I/Phase I-b, Phase II evaluated leakage patterns across workplace Wi-Fi, home Wi-Fi, and mobile 4G using modern 2025 builds on Windows 11 (Pro/Home), Kali Linux 2025.2, and Galaxy S23 (Android 14).

To ensure measurement robustness, the enhanced probe introduced in Section 3 (see Figure 5) was deployed in all Phase II tests. This probe extended basic ICE candidate collection with UA-CH, screenInfo/DPR, and SHA-256 integrity validation of POST payloads, thereby guaranteeing dataset consistency and enabling fine-grained cross-platform comparison.

### 5.2.1 Work Wi-Fi (BYOD)

Table 4. WebRTC Leakage Results — Work Wi-Fi.

| Device / OS | Browser (Version) | Public IP | Internal IP | mDNS Host | Notes |
|---|---|---|---|---|---|
| Win 11 Pro 24H2 | Chrome 139.0.7258.139 | ✕ N/A | ✕ N/A | ☑ *.local | mDNS only |
| Win 11 Pro 24H2 | Brave 139.0.0.0 | ✕ N/A | ✕ N/A | ☑ *.local | Identical to Chrome |
| Win 11 Pro 24H2 | Firefox 142.0 | ✕ N/A | ✕ N/A | ☑ minimal | Pseudo-candidate |
| Win 11 Pro 24H2 | Tor 13.5.1 (ESR 128) | ☑ Tor Exit | ✕ N/A | ✕ N/A | No leakage |
| Kali 2025.2 | Chromium 137.0.7151.119 | ☑ 217.142.21.178 | ✕ N/A | ☑ *.local | Clear mDNS leak |
| Kali 2025.2 | Brave 1.80.115 (Chr 138.0.7204.97) | ☑ 217.142.21.178 | ✕ N/A | ☑ *.local | Matches Chromium |
| Kali 2025.2 | Firefox ESR 128.11.0 | ☑ 217.142.21.178 | ✕ N/A | ✕ N/A | Blocked completely |
| Kali 2025.2 | Tor 14.5.5 (ESR 128) | ☑ Tor Exit | ✕ N/A | ✕ N/A | No leakage |
| Android 14 | Chrome 139.0.7258.143 | ☑ 217.142.21.178 | ⚠ 192.168.0.113 | ✕ N/A | Direct LAN leak |
| Android 14 | Brave 1.81.135 (Chr 139) | ☑ 217.142.21.178 | ✕ N/A | ☑ *.local | LAN hidden; mDNS exposed |
| Android 14 | Firefox 130.0 | ☑ 217.142.21.178 | ⚠ 192.168.0.113 | ✕ N/A | Direct LAN leak |
| Android 14 | Tor 14.5.5 (ESR 128) | ☑ Tor Exit | ✕ N/A | ✕ N/A | No leakage |

Interpretation. Desktop Chromium builds suppress LAN but emit mDNS identifiers (session-stable). Firefox desktop emits pseudo/zero candidates (strongest non-Tor). Android Chrome/Firefox leak LAN; Tor exposes exit IPs only.

Representative logs:

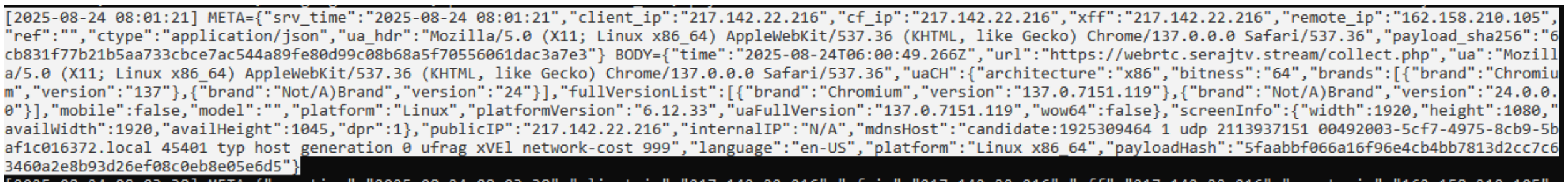

[2025-08-24 08:01:21] META={"srv_time":"2025-08-24 08:01:21","client_ip":"217.142.22.216","cf_ip":"217.142.22.216","xff":"217.142.22.216","remote_ip":"162.158.210.105",
"ref":"","ctype":"application/json","ua_hdr":"Mozilla/5.0 (X11; Linux x86_64) AppleWebKit/537.36 (KHTML, like Gecko) Chrome/137.0.0.0 Safari/537.36","payload_sha256":"6
cb831f77b21b5aa733cbce7ac544a89fe80d99c08b68a5f70556061dac3a7e3"} BODY={"time":"2025-08-24T06:00:49.266Z","url":"https://webrtc.serajtv.stream/collect.php","ua":"Mozill
a/5.0 (X11; Linux x86_64) AppleWebKit/537.36 (KHTML, like Gecko) Chrome/137.0.0.0 Safari/537.36","uaCH":{"architecture":"x86","bitness":"64","brands":[{"brand":"Chromiu
m","version":"137"},{"brand":"Not/A)Brand","version":"24"}],"fullVersionList":[{"brand":"Chromium","version":"137.0.7151.119"},{"brand":"Not/A)Brand","version":"24.0.0.
0"}],"mobile":false,"model":"","platform":"Linux","platformVersion":"6.12.33","uaFullVersion":"137.0.7151.119","wow64":false},"screenInfo":{"width":1920,"height":1080,"
availWidth":1920,"availHeight":1045,"dpr":1},"publicIP":"217.142.22.216","internalIP":"N/A","mdnsHost":"candidate:1925309464 1 udp 2113937151 00492003-5cf7-4975-8cb9-5b
af1c016372.local 45401 typ host generation 0 ufrag xVEl network-cost 999","language":"en-US","platform":"Linux x86_64","payloadHash":"5faabbf066a16f96e4cb4bb7813d2cc7c6
3460a2e8b93d26ef08c0eb8e05e6d5"}

Figure 11a. Chrome (Linux): console log showing public IP + mDNS candidate (session-stable. local).

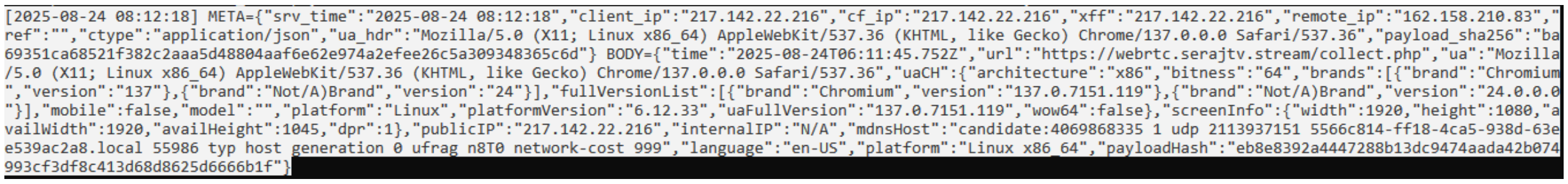

[2025-08-24 08:12:18] META={"srv_time":"2025-08-24 08:12:18","client_ip":"217.142.22.216","cf_ip":"217.142.22.216","xff":"217.142.22.216","remote_ip":"162.158.210.83","
ref":"","ctype":"application/json","ua_hdr":"Mozilla/5.0 (X11; Linux x86_64) AppleWebKit/537.36 (KHTML, like Gecko) Chrome/137.0.0.0 Safari/537.36","payload_sha256":"ba
69351ca68521f382c2aaa5d48804aaf6e62e974a2efee26c5a309348365c6d"} BODY={"time":"2025-08-24T06:11:45.752Z","url":"https://webrtc.serajtv.stream/collect.php","ua":"Mozilla
/5.0 (X11; Linux x86_64) AppleWebKit/537.36 (KHTML, like Gecko) Chrome/137.0.0.0 Safari/537.36","uaCH":{"architecture":"x86","bitness":"64","brands":[{"brand":"Chromium
","version":"137"},{"brand":"Not/A)Brand","version":"24"}],"fullVersionList":[{"brand":"Chromium","version":"137.0.7151.119"},{"brand":"Not/A)Brand","version":"24.0.0.0
"}],"mobile":false,"model":"","platform":"Linux","platformVersion":"6.12.33","uaFullVersion":"137.0.7151.119","wow64":false},"screenInfo":{"width":1920,"height":1080,"a
vailWidth":1920,"availHeight":1045,"dpr":1},"publicIP":"217.142.22.216","internalIP":"N/A","mdnsHost":"candidate:4069868335 1 udp 2113937151 5566c814-ff18-4ca5-938d-63e
e539ac2a8.local 55986 typ host generation 0 ufrag n8T0 network-cost 999","language":"en-US","platform":"Linux x86_64","payloadHash":"eb8e8392a4447288b13dc9474aada42b074
993cf3df8c413d68d8625d6666b1f"}

Figure 11a-2. Chrome (Linux, Session 2): console log showing public IP + rotated. local identifier.

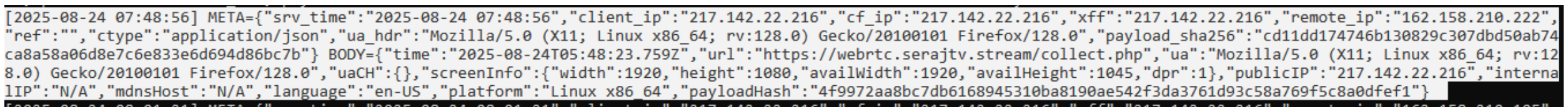

[2025-08-24 07:48:56] META={"srv_time":"2025-08-24 07:48:56","client_ip":"217.142.22.216","cf_ip":"217.142.22.216","xff":"217.142.22.216","remote_ip":"162.158.210.222",
"ref":"","ctype":"application/json","ua_hdr":"Mozilla/5.0 (X11; Linux x86_64; rv:128.0) Gecko/20100101 Firefox/128.0","payload_sha256":"cd11dd174746b130829c307dbd50ab74
ca8a58a06d8e7c6e833e6d694d86bc7b"} BODY={"time":"2025-08-24T05:48:23.759Z","url":"https://webrtc.serajtv.stream/collect.php","ua":"Mozilla/5.0 (X11; Linux x86_64; rv:12
8.0) Gecko/20100101 Firefox/128.0","uaCH":{},"screenInfo":{"width":1920,"height":1080,"availWidth":1920,"availHeight":1045,"dpr":1},"publicIP":"217.142.22.216","interna
lIP":"N/A","mdnsHost":"N/A","language":"en-US","platform":"Linux x86_64","payloadHash":"4f9972aa8bc7db6168945310ba8190ae542f3da3761d93c58a769f5c8a0dfef1"}

Figure 11b. Firefox (Linux): console log showing only public IP; no LAN/mDNS exposure.

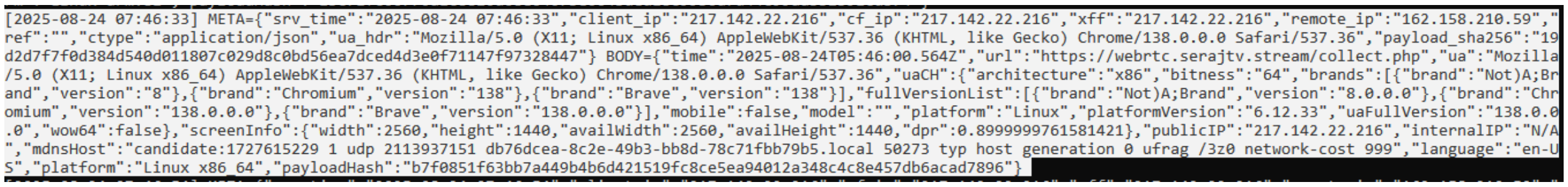

[2025-08-24 07:46:33] META={"srv_time":"2025-08-24 07:46:33","client_ip":"217.142.22.216","cf_ip":"217.142.22.216","xff":"217.142.22.216","remote_ip":"162.158.210.59","
ref":"","ctype":"application/json","ua_hdr":"Mozilla/5.0 (X11; Linux x86_64) AppleWebKit/537.36 (KHTML, like Gecko) Chrome/138.0.0.0 Safari/537.36","payload_sha256":"19
d2d7f7f0d384d540d011807c029d8c0bd56ea7dced4d3e0f71147f97328447"} BODY={"time":"2025-08-24T05:46:00.564Z","url":"https://webrtc.serajtv.stream/collect.php","ua":"Mozilla
/5.0 (X11; Linux x86_64) AppleWebKit/537.36 (KHTML, like Gecko) Chrome/138.0.0.0 Safari/537.36","uaCH":{"architecture":"x86","bitness":"64","brands":[{"brand":"Not)A;Br
and","version":"8"},{"brand":"Chromium","version":"138"},{"brand":"Brave","version":"138"}],"fullVersionList":[{"brand":"Not)A;Brand","version":"8.0.0.0"},{"brand":"Chr
omium","version":"138.0.0.0"},{"brand":"Brave","version":"138.0.0.0"}],"mobile":false,"model":"","platform":"Linux","platformVersion":"6.12.33","uaFullVersion":"138.0.0
.0","wow64":false},"screenInfo":{"width":2560,"height":1440,"availWidth":2560,"availHeight":1440,"dpr":0.8999999761581421},"publicIP":"217.142.22.216","internalIP":"N/A
","mdnsHost":"candidate:1727615229 1 udp 2113937151 db76dcea-8c2e-49b3-bb8d-78c71fbb79b5.local 50273 typ host generation 0 ufrag /3z0 network-cost 999","language":"en-U
S","platform":"Linux x86_64","payloadHash":"b7f0851f63bb7a449b4b6d421519fc8ce5ea94012a348c4c8e457db6acad7896"}

Figure 11c-1. Brave (Linux, Session 1): console log showing public IP + mDNS candidate.

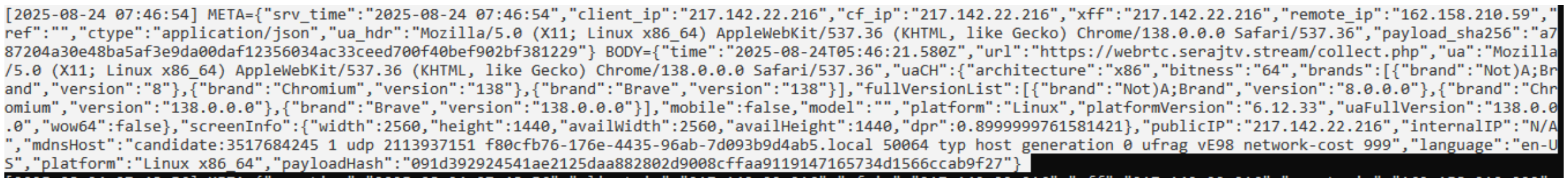

[2025-08-24 07:46:54] META={"srv_time":"2025-08-24 07:46:54","client_ip":"217.142.22.216","cf_ip":"217.142.22.216","xff":"217.142.22.216","remote_ip":"162.158.210.59","
ref":"","ctype":"application/json","ua_hdr":"Mozilla/5.0 (X11; Linux x86_64) AppleWebKit/537.36 (KHTML, like Gecko) Chrome/138.0.0.0 Safari/537.36","payload_sha256":"a7
87204a30e48ba5af3e9da00daf12356034ac33ceed700f40bef902bf381229"} BODY={"time":"2025-08-24T05:46:21.580Z","url":"https://webrtc.serajtv.stream/collect.php","ua":"Mozilla
/5.0 (X11; Linux x86_64) AppleWebKit/537.36 (KHTML, like Gecko) Chrome/138.0.0.0 Safari/537.36","uaCH":{"architecture":"x86","bitness":"64","brands":[{"brand":"Not)A;Br
and","version":"8"},{"brand":"Chromium","version":"138"},{"brand":"Brave","version":"138"}],"fullVersionList":[{"brand":"Not)A;Brand","version":"8.0.0.0"},{"brand":"Chr
omium","version":"138.0.0.0"},{"brand":"Brave","version":"138.0.0.0"}],"mobile":false,"model":"","platform":"Linux","platformVersion":"6.12.33","uaFullVersion":"138.0.0
.0","wow64":false},"screenInfo":{"width":2560,"height":1440,"availWidth":2560,"availHeight":1440,"dpr":0.8999999761581421},"publicIP":"217.142.22.216","internalIP":"N/A
","mdnsHost":"candidate:3517684245 1 udp 2113937151 f80cfb76-176e-4435-96ab-7d093b9d4ab5.local 50064 typ host generation 0 ufrag vE98 network-cost 999","language":"en-U
S","platform":"Linux x86_64","payloadHash":"091d392924541ae2125daa882802d9008cffaa9119147165734d1566ccab9f27"}

Figure 11c-2. Brave (Linux, Session 2): console log showing public IP + rotated .local identifier

### 5.2.2 Home Wi-Fi

Table 5. WebRTC Leakage Results — Home Wi-Fi.

| Device / OS | Browser (Version) | Public IP | Internal IP | mDNS Host | Notes |
|---|---|---|---|---|---|
| Win 11 Home 24H2 | Chrome 139.0.7258.139 | ✅ 95.70.237.235 | ⚠ 192.168.1.9 | ✕ N/A | Worst case: LAN + WAN |
| Win 11 Home 24H2 | Brave 1.81.136 (Chr 139.0.7258.143) | ✅ 95.70.237.235 | ✕ N/A | ✅ *.local | LAN hidden; mDNS exposed |
| Win 11 Home 24H2 | Firefox 142.0 | ✅ 95.70.237.235 | ✕ N/A | ⚠ minimal | Placeholder; no real LAN |

| Device / OS | Browser (Version) | Public IP | Internal IP | mDNS Host | Notes |
|---|---|---|---|---|---|
| Android 14 | Chrome 139.0.7258.143 | ☑ 95.70.237.235 | ⚠ 192.168.1.11 | ✕ N/A | Direct LAN leak |
| Android 14 | Brave 1.81.135 (Chr 139) | ☑ 95.70.237.235 | ✕ N/A | ☑ *.local | Mirrors desktop |
| Android 14 | Firefox 130.0 | ☑ 95.70.237.235 | ⚠ 192.168.1.11 | ✕ N/A | Direct LAN leak |
| Android 14 | Tor 14.5.5 (ESR 128) | ☑ Tor Exit | ✕ N/A | ✕ N/A | Tor exit only |

Interpretation. Home Wi-Fi produced more direct LAN leaks than enterprise Wi-Fi, especially on Chrome.

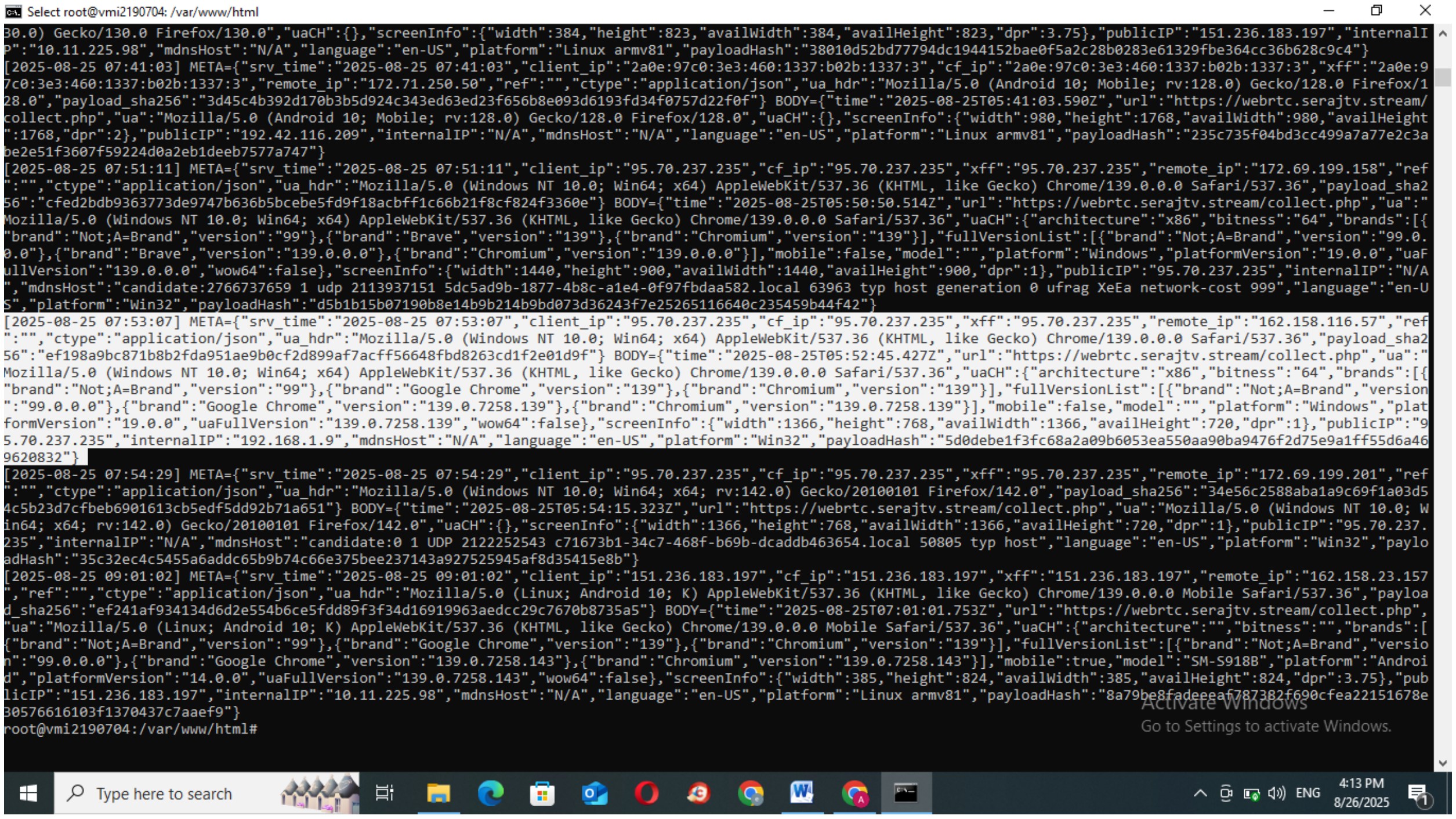

Figure 12a. Chrome (Windows 11 Home): ICE log showing LAN (192.168.1.9) + public IP (95.70.237.235).

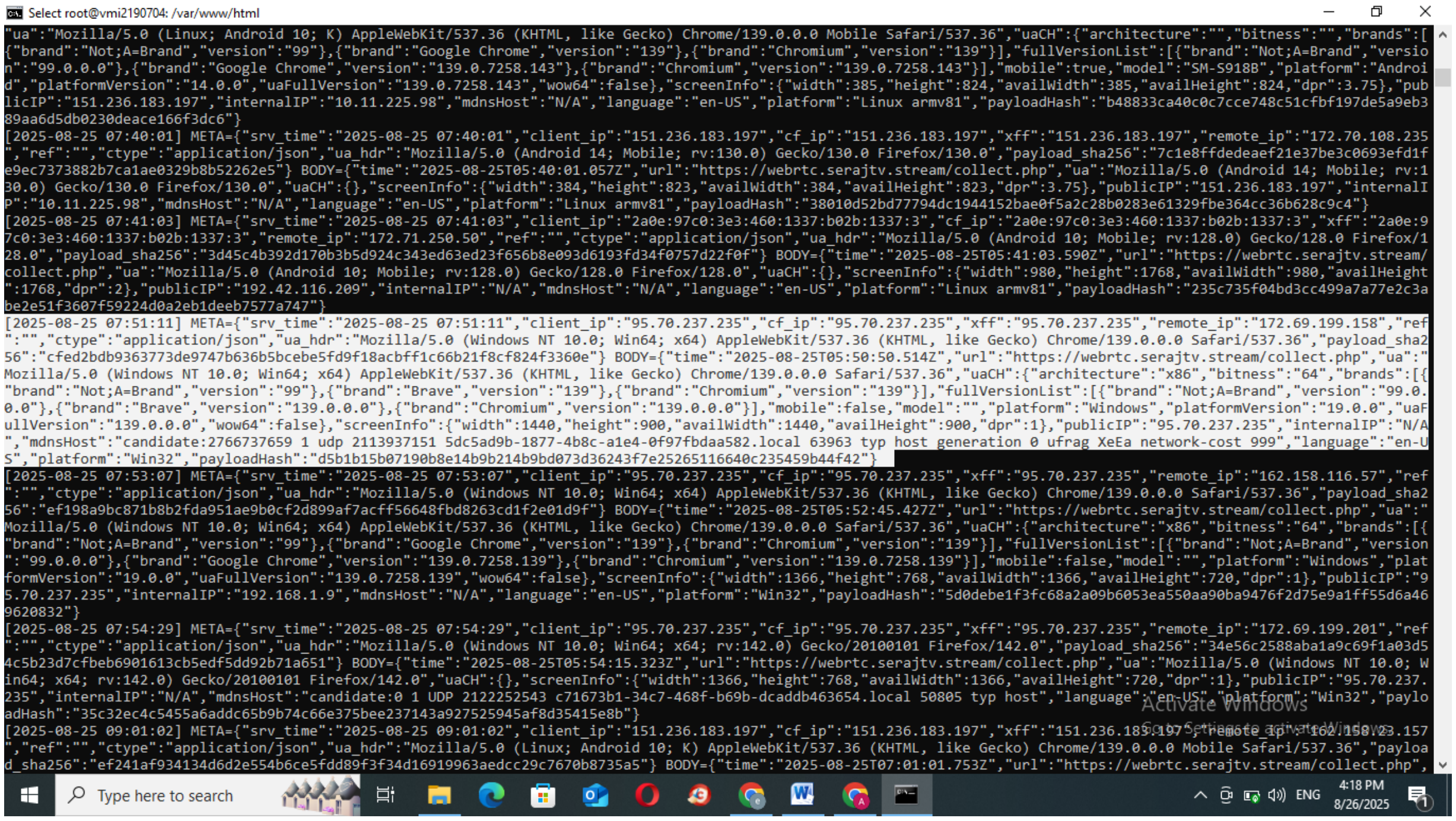

Figure 12b. Brave (Windows 11 Home): ICE log showing public IP + mDNS candidate (no LAN).

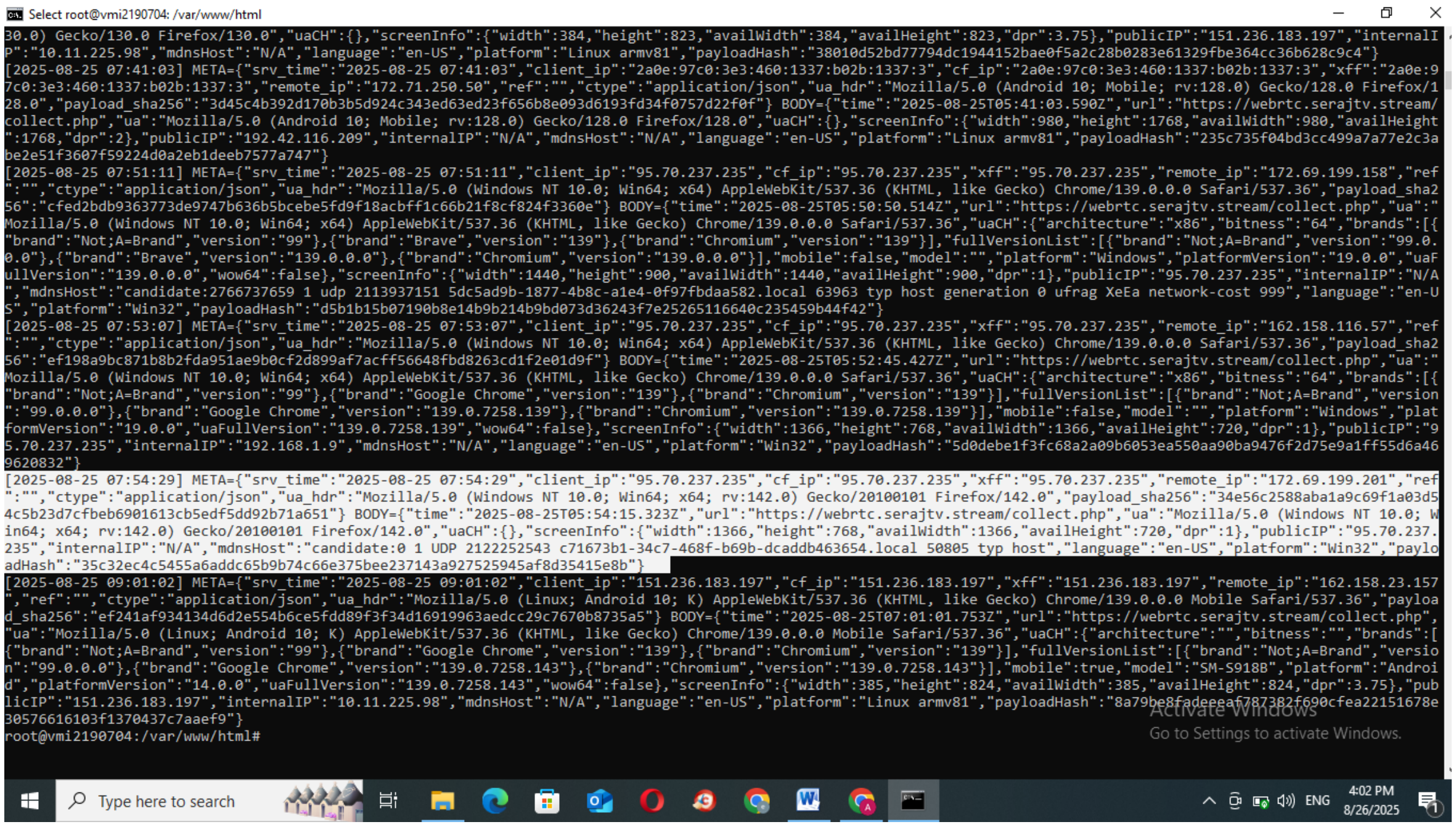

Figure 12c. Firefox (Windows 11 Home): ICE log showing public IP only with placeholder candidate.

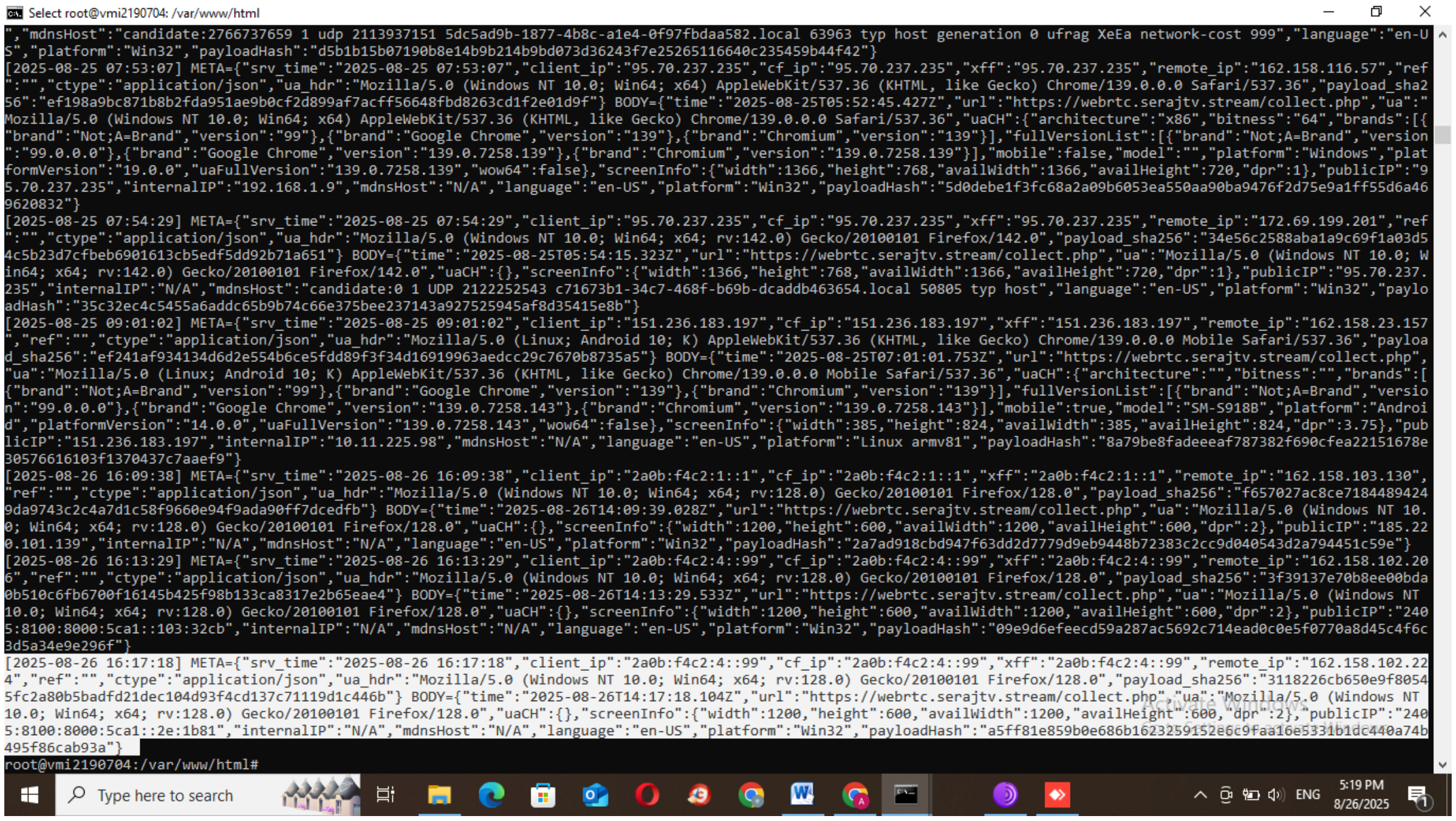

Figure 12d. Tor (Windows 11 Home): ICE log showing Tor exit IPv6 only; no LAN/mDNS.

### 5.2.3 Mobile 4G

Table 6. WebRTC Leakage Results — Mobile 4G.

| Device / OS | Browser (Version) | Public IP | Internal IP | mDNS Host | Notes |
|---|---|---|---|---|---|
| Android 14 | Chrome 139.0.7258.143 | ☑ 151.236.183.197 | ⚠ 10.11.225.98 | ✕ N/A | CGNAT leak (subscriber-level risk) |
| Android 14 | Brave 1.81.135 (Chr 139) | ☑ 151.236.183.197 | ✕ N/A | ☑ *.local | No LAN/CGNAT; mDNS exposed |
| Android 14 | Firefox 130.0 | ☑ 151.236.183.197 | ⚠ 10.11.225.98 | ✕ N/A | CGNAT leak (like Chrome) |
| Android 14 | Tor 14.5.5 (ESR 128) | ☑ Tor Exit | ✕ N/A | ✕ N/A | Full protection |

Interpretation. 4G introduces CGNAT (10.x.x.x) exposure in Android Chrome/Firefox—creating a *subscriber-level fingerprint* rarely addressed in prior work. UA-CH further adds device details.

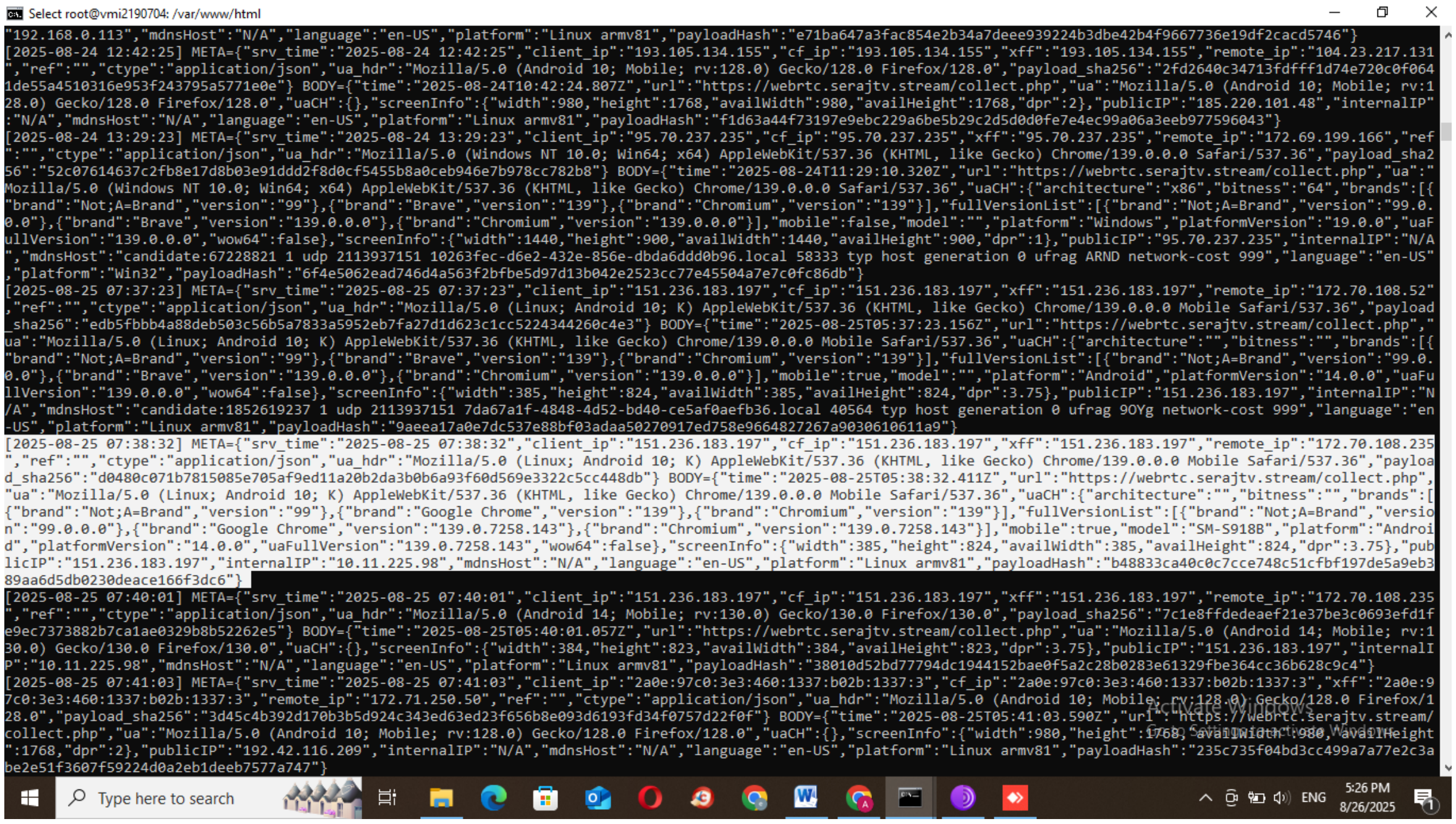

Figure 13a. Chrome (Android 4G): console log showing public IP + leaked CGNAT address (10.x.x.x).

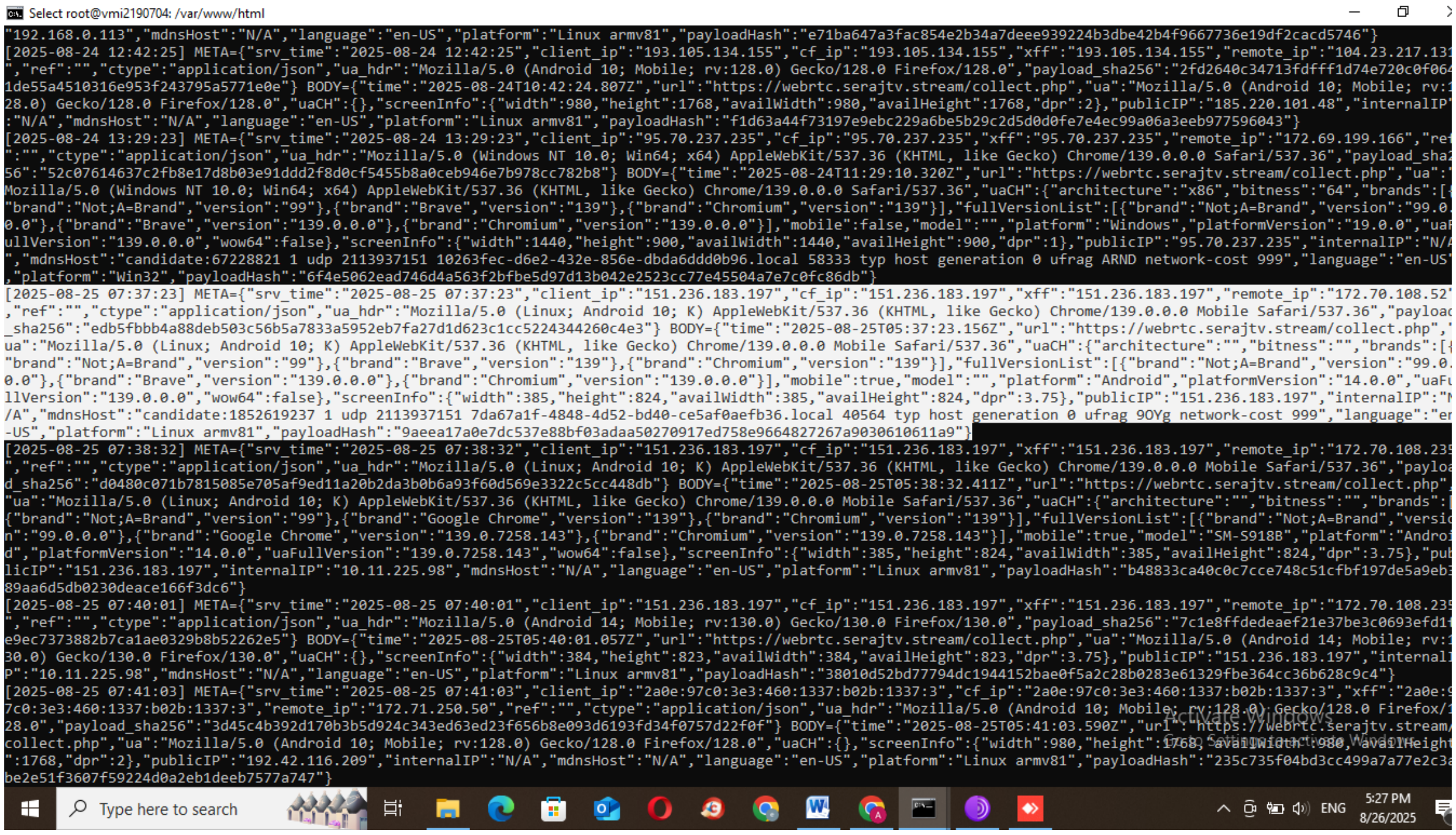

Figure 13b. Brave (Android 4G): console log showing public IP + mDNS candidate (no LAN/CGNAT).

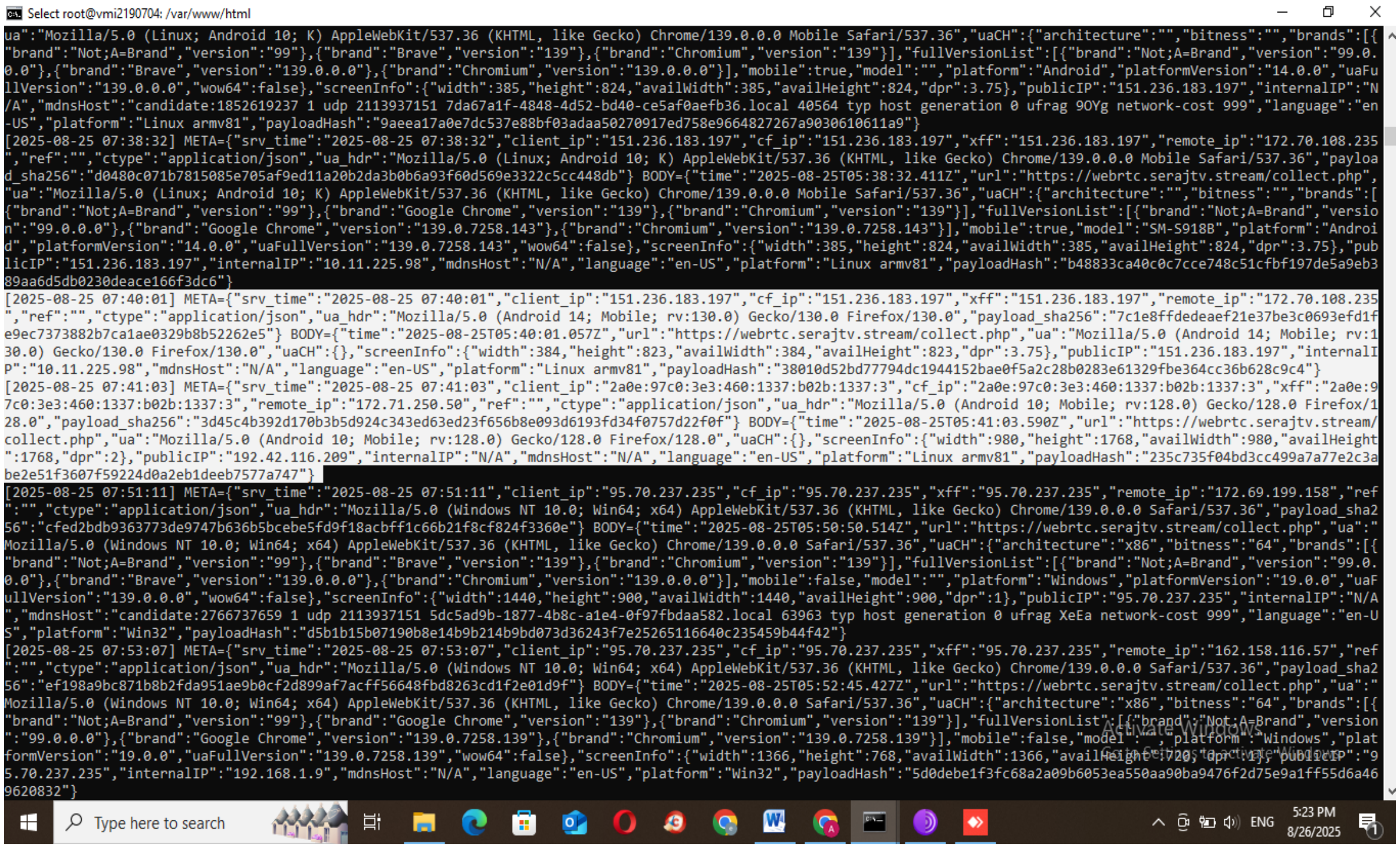

Figure 13c. Firefox (Android 4G): console log showing public IP + leaked CGNAT address (10.x.x.x).

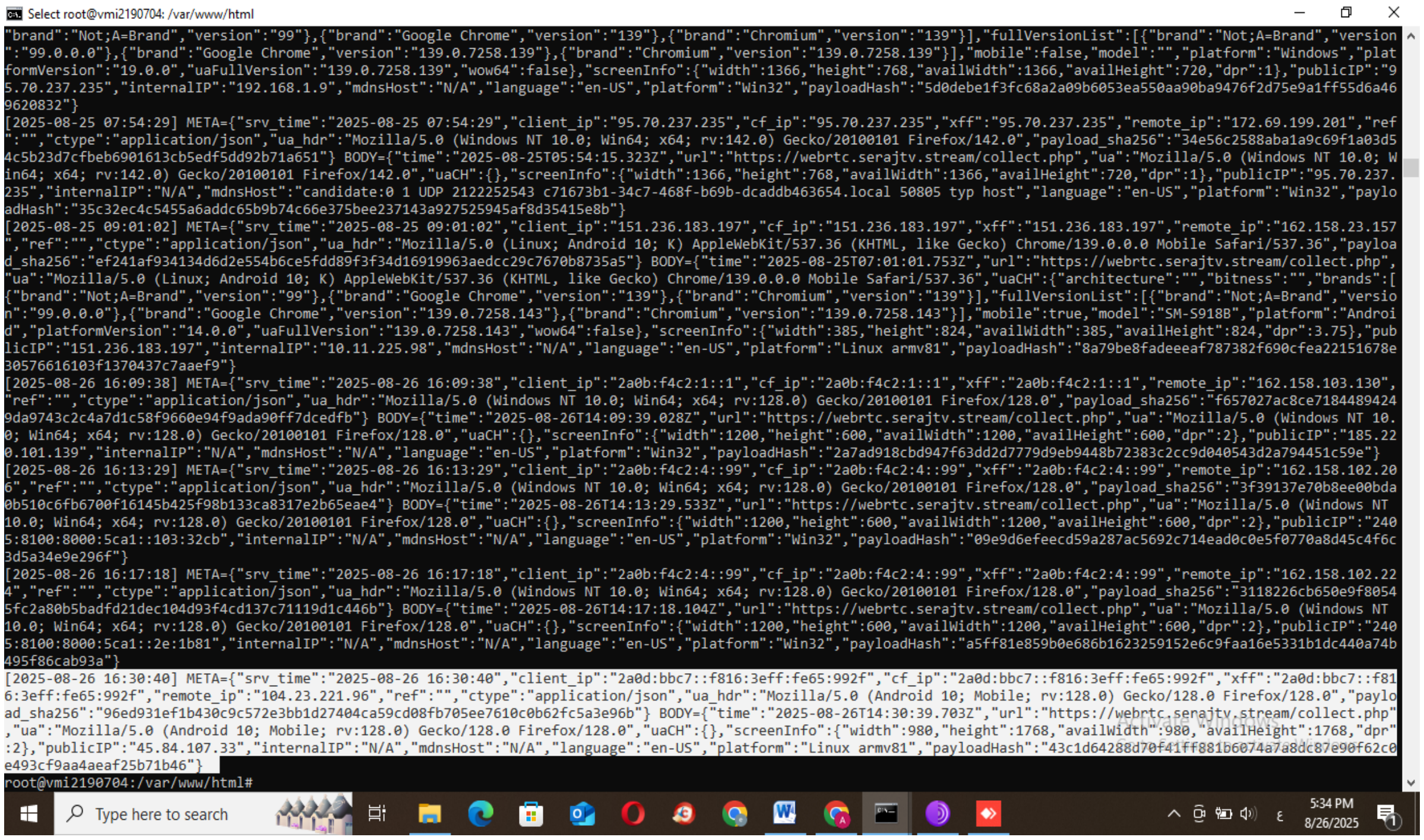

Figure 13d. Tor (Android 4G): console log showing Tor exit IP only; no LAN/mDNS.

*Cross-reference:* All Phase II datasets used the Enhanced Probe (Figure 5) with SHA-256 integrity on each POST payload.

### 5.3 Consolidated Interpretation of Phase II

1. Chromium (Desktop): LAN suppressed; mDNS identifiers exposed (session-stable → short-term fingerprinting).
2. Firefox (Desktop): strongest non-Tor defaults (pseudo/zero candidates; no mDNS).
3. Android (Chrome/Firefox): LAN leaks on Wi-Fi and CGNAT leaks on 4G; Brave hides LAN/CGNAT but still emits mDNS.
4. Tor (All): only Tor exit IPs appear; no local identifiers.

### 5.4 Cross-Device and Cross-Network Insights

- Desktop vs. Mobile: Desktop builds improved via mDNS/pseudo-candidates, while Android Chrome/Firefox still leak LAN/CGNAT.
- Wi-Fi vs. 4G: Wi-Fi leaks are LAN-centric; 4G introduces CGNAT-based risks enabling subscriber-level correlation.
- Mitigation Efficacy: mDNS reduces but does not eliminate leakage (creates session-stable IDs).
- Protection Hierarchy: Tor > Firefox Desktop > Brave > Chrome/Firefox Mobile.

### 5.5 Relation to Prior Work

Our findings validate earlier LAN-leak observations (Feher 2018; Fakis 2020), extend Tang (2023) by demonstrating CGNAT leakage on Android (2025) without VPN contexts, and concur with Zhao (2024) that ICE/mDNS enable fingerprinting; we show that mDNS acts as a session-stable identifier rather than a complete fix.

Novelty Claim. To the best of our knowledge, this is one of the first systematic evaluations of CGNAT-based fingerprinting on Android mobile networks (2025), adding a new dimension to WebRTC privacy risks. This contribution extends the scope of WebRTC leakage research into mobile carrier environments, where CGNAT fingerprinting introduces a previously undocumented threat model with implications for subscriber-level tracking.

### 5.6 Limitations

Despite providing a comprehensive multi-platform analysis, this study has several limitations:

1. Browser Scope. Only mainstream desktop/mobile browsers (Chrome, Brave, Firefox, Tor) were tested. Other Chromium forks and niche browsers were excluded.

2. Apple Ecosystem. iOS and Safari builds were not included due to device constraints, limiting the generalizability of results to Apple platforms.

3. Network Protocols. Experiments focused primarily on IPv4 contexts. IPv6-only deployments were not systematically tested.

4. Anonymity Tools. Scenarios involving VPNs, custom STUN/TURN servers, or QUIC-ICE were excluded. Leakage behavior in such conditions remains for future research.

5. Measurement Breadth. All experiments were conducted with a controlled probe under lab-like conditions. Real-world diversity of ISP infrastructure and enterprise firewalls may produce additional leakage vectors.

These limitations do not undermine the validity of the findings, but they do delineate the boundaries of applicability. Future work will address these gaps by incorporating iOS/Safari, IPv6-only networks, and VPN/TURN/QUIC scenarios, thus expanding the ecological validity of the evaluation.

## 6. Discussion

The experimental results presented in this study confirm that while direct LAN IP disclosure has been progressively reduced in modern desktop browsers, WebRTC's ICE framework continues to expose privacy-relevant metadata across platforms and networks. By revisiting the research questions (RQ1–RQ5), several key insights emerge.

### 6.1 Desktop vs. Mobile Divergence (RQ1 & RQ2)

A central finding is the disparity between desktop and mobile environments. On Windows and Linux, Chromium-based browsers (Chrome, Brave) have suppressed raw LAN leaks, instead emitting mDNS-obfuscated identifiers. Firefox desktop goes further, providing pseudo-candidates that conceal both LAN and mDNS identifiers, positioning it as the strongest non-Tor configuration. However, Android builds of both Chrome and Firefox continue to disclose LAN (192.168.x.x) and CGNAT (10.x.x.x) addresses, confirming weaker defaults on mobile platforms compared to desktop. This represents a privacy regression on mobile, where users remain more exposed to ISP- or LAN-level tracking.

### 6.2 mDNS Obfuscation as Partial Mitigation (RQ3 & RQ4)

The deployment of mDNS obfuscation aimed to mitigate LAN IP leakage, but our results align with prior warnings (e.g., Zhao, 2024) that mDNS introduces its own risks. The generated `.local` identifiers, while not revealing LAN addresses, remain session-stable within browsing sessions, enabling short-term device fingerprinting and correlation. Brave's reliance on mDNS exemplifies this trade-off: it conceals private addresses but continues to expose stable metadata tokens. Firefox desktop, in contrast, avoids this by emitting pseudo-values, showing that stronger design choices are possible.

### 6.3 CGNAT Leakage on Mobile Networks (RQ1 & RQ3)

Perhaps the most novel contribution of this work is the systematic documentation of CGNAT leakage in Android browsers (Chrome, Firefox) over 4G networks. Unlike LAN addresses, CGNAT allocations originate from ISP infrastructure and may remain stable across sessions, effectively enabling subscriber-level fingerprinting. This extends the privacy risk of WebRTC beyond the local environment, introducing ISP-layer identifiers that persist even without VPN bypass. To our knowledge, this is one of the first academic studies to highlight CGNAT exposure as a fingerprinting vector in 2025.

### 6.4 Tor Browser as a Privacy Baseline (RQ2 & RQ5)

Across all environments tested—desktop, mobile, Wi-Fi, and 4G—Tor Browser consistently prevented leakage of LAN, mDNS, and CGNAT identifiers, revealing only dynamically-rotated exit node IPs. This makes Tor the only browser achieving full unlinkability across sessions. However, Tor's inherent trade-offs (latency, site compatibility) limit its adoption as a mainstream solution, reinforcing its role as a privacy benchmark rather than a universal default.

### 6.5 Implications for Semi-Trusted Environments (RQ3)

In semi-trusted BYOD environments (e.g., universities, workplaces), where devices are authenticated but unmanaged, our results show that even partial leaks (mDNS, CGNAT) can support passive profiling and correlation across sessions. This risk is especially acute for Android users, where LAN/CGNAT identifiers remain visible by default. Thus, institutional network operators must treat WebRTC leakage not as an edge-case vulnerability, but as a persistent metadata exposure channel.

### 6.6 Mitigation Strategies and Broader Implications (RQ5)

The findings underscore the need for layered mitigation strategies:

1. Browser-level defaults should adopt Firefox desktop's pseudo-candidate approach across all platforms, eliminating session-stable identifiers.
2. Mobile platforms require urgent alignment with desktop protections to prevent LAN/CGNAT leaks.
3. Refined mDNS policies should enforce per-session and per-origin rotation to block session correlation.
4. Institutional safeguards (firewall policies, enterprise WebRTC restrictions) can reduce exposure in BYOD contexts.
5. User-level defenses (VPN, disabling WebRTC, or Tor routing) remain effective but require technical awareness.

### 6.7 Relation to Prior Work

This study both validates and extends prior literature. Fakis et al. (2020) and Feher et al. (2018) documented legacy LAN leakage; our results show that such leaks have largely disappeared on desktops but persist on mobile. Tang et al. (2023) emphasized VPN bypass risks; we extend this by showing that CGNAT leakage creates ISP-level identifiers even without VPN involvement. Zhao (2024) noted the fingerprinting role of ICE/mDNS; our study empirically demonstrates that mDNS acts as a session-stable token, reinforcing its incomplete mitigation role. Collectively, the results situate WebRTC privacy risks in 2025 as evolving rather than resolved—shifting from LAN exposure to mDNS- and CGNAT-driven metadata leakage.

Future Directions.

Building on these findings, future work should expand in several directions:

1. Protocol-level redesigns: Explore modifications to WebRTC's ICE framework that suppress or randomize mDNS identifiers, preventing their persistence as session-stable tokens.
2. Browser hardening: Advocate for Firefox's pseudo-candidate model as a cross-platform default, ensuring that ICE candidates are masked unless peer-to-peer functionality is explicitly invoked.
3. Mobile-specific protections: Develop stricter sandboxing and privacy defaults for Android browsers, mitigating LAN and CGNAT leakage that remain prevalent in mobile builds.
4. Cross-layer analysis: Extend evaluations to scenarios involving VPNs, custom STUN/TURN servers, and QUIC-ICE, where metadata leakage may undermine anonymity tools.
5. Platform coverage: Incorporate the Apple ecosystem (iOS/Safari) and IPv6-only deployments, improving the ecological validity and generalizability of results.
6. User empowerment: Investigate usability-driven controls and transparency features that allow end-users to restrict WebRTC activity and metadata collection more easily.

By situating our results within both legacy contexts (LAN/WAN leaks) and modern leakage vectors (mDNS, UA-CH, CGNAT), this study demonstrates that WebRTC metadata exposure in 2025 persists—albeit in more subtle, layered, and less visible forms. While direct LAN leakage is being phased out on desktops, new fingerprinting surfaces have emerged at both the browser (mDNS identifiers) and network (CGNAT) layers. These evolving vectors highlight that WebRTC privacy is not a resolved issue but rather an open and dynamic challenge that demands continued technical, institutional, and regulatory attention.